\documentclass[article]{revtex4}
\usepackage{graphicx}
\usepackage[utf8]{inputenc}
\usepackage{dcolumn}
\usepackage{placeins}
\usepackage{amsmath}
\usepackage{siunitx}
\usepackage{caption}
\usepackage{adjustbox}
\usepackage[colorlinks=true, citecolor=blue, urlcolor = blue, linkcolor= red, bookmarks=true]{hyperref}
\usepackage{float}
\usepackage{tabularray}
\usepackage{adjustbox}
\usepackage{amsfonts}
\usepackage{dcolumn}
\usepackage{float}
\usepackage{hyperref}
\usepackage{subcaption}
\usepackage{epstopdf}
\usepackage{booktabs}
\usepackage{multirow}
\usepackage{bm}

\makeatletter
\def\btt#1{\texttt{\@backslashchar#1}}
\DeclareRobustCommand\bblash{\btt{\@backslashchar}} \makeatother

\makeatletter
\def\btt#1{\texttt{\@backslashchar#1}}
\DeclareRobustCommand\bblash{\btt{\@backslashchar}} \makeatother

\begin{document}
\title{Analysis of Solar Flare and Sunspots on 4th Jan 2025 and Their Effects on Space Weather }
\author{Akash Vinod Shirke $^{a}$}\email{Akash.11462@stu.upes.ac.in}
\author{Sakshi Sheshrao Charde $^{a}$}\email{Sakshi.11314@stu.upes.ac.in}
\author{Balendra Pratap Singh$^{a}$}\email{balendrap.singh@ddn.upes.ac.in}
\affiliation{ $^{a}$ Department of Physics, Applied Science Cluster, SOAE, UPES, Energy Acres, Bidholi, Via Prem Nagar,  Dehradun, Uttarakhand, 248007, India}
\begin{abstract}
Solar flares and coronal mass ejections (CMEs) are among the most energetic
phenomena in the solar system, often impacting space weather and terrestrial technologies. In this study, we utilize SunPy, an open-source Python library for solar
physics, to analyze solar active regions and their correlation with flare and CME
events observed on 4th January 2025. Data from GOES, SDO (AIA and HMI), Solar Oribter (STIX), e-CALLISTO, Aditya L1 (SUIT), and
SOHO are processed to track flare intensity, active region evolution, shock wave and CME
dynamics. The analyzed flare is identified as an X1.8-class event, and our study
highlights key magnetic precursors that led to it. This work enhances understanding of solar eruption precursors and supports future predictive models for space
weather forecasting.
\\
\noindent\textbf{Keywords:} SunPy; Solar flares; Coronal mass ejections; Sunspots; Magnetic field structure; Solar active regions.
\end{abstract}
\maketitle
\section{Introduction}
The intense magnetic dynamics of the Sun results in observable signatures appear as the solar flares and sunspots, typically driving the space weather events that can majorly affect communication systems, satellites, and power grids \cite{Schrijver2010}. These variable events arise around a magnetic complex active regions and regulated by the long-term solar cycle, which also impacts their intensity and frequencies \cite{Hathaway2015}. Understanding the origin, temporal variation, and consequences of such eruptive events is essential for improving space weather forecasting. The abrupt burst of electromagnetic radiation is defined as Solar Flares, further releasing the magnetic free energy into the solar atmosphere, resulting from magnetic reconnection mechanism in active region \cite{Shibata2011}. Especially the powerful X-class Solar flares can emit around $10^{31}$–$10^{32}$ erg in radiation and often linked with coronal mass ejections (CMEs), radio blackouts, and geomagnetic storms \cite{Emslie2012,Hudson2011}. From massive, abruptly evolving sunspot groups, these types of intense flares occur, which are continuously tracked and classified by agencies like NOAA \cite{Sammis2000}. During the solar maximum period, the frequency of these types of events peaks rapidly, for the current Solar Cycle-25 which was started in late 2019 and the predicted maximum sunspot number ranges around 105 to 125, predicted to peak nearly mid-2025 \cite{NOAA2020}. Consistent with this forecast, early 2025 has already observed heightening solar activity, with some several X-class flares detected from NOAA Active Region (AR) 13947. On 4 January 2025, this active region produced a major X1.8-class solar flare, which peaked around 12:47 UTC, following a X1.1 and multiple M-class solar flares detected the previous day \cite{NOAA2025}. In this paper, we present a source to eruption detailed analysis by data-driven investigation of January 4, 2025 X1.8 -class flare. Sect.~(\ref{sect2}) describes the data sources and analysis methods, including the SunPy-based routines \cite{2020ApJ...890...68S}. Sect.~(\ref{sect3}) details our findings, mainly covering the flare energetics, reconnection measures, active region classification and linked with a CME. In Sect.~(\ref{sect4}), we discussed the broader consequences and space weather effects.
\section{Data and Methodology}\label{sect2}
We used a combination of satellite observations and analysis tools. GOES X-ray data: Geostationary Operational Environmental Satellites (GOES) carry an X-Ray Sensor (XRS)  that measures solar soft X-rays in the band, ranging from 0.5-4 and 1-8. The GOES 1-8 flux defines flare class (e.g., X1.8 = $1.8\times10^{-4}$ W / m$^2$ peak). We downloaded GOES XRS data of January 4, 2025 using the \texttt{Fido} data access client in SunPy and constructed a time series of the soft X-ray flux, identifying a peak at 12:47(UTC) \cite{GOESXRS}. This yields the flare light curve $F(t)$, from which we estimate the peak flux. SDO/AIA imaging: We retrieved extreme-ultraviolet (EUV) images from the Atmospheric Imaging Assembly (AIA) onboard the Solar Dynamics Observatory (SDO) in the 131, 171, 193, and 304~\AA{} channels. The data were downloaded using SunPy's \texttt{Fido} client for 4--5 January 2025 to visualize the flare and filament activity \cite{2020ApJ...890...68S}. The AIA 131~\AA{} channel is sensitive to hot plasma at temperatures of $\sim$10--15~MK, typically associated with solar flares, whereas the 304~\AA{} channel primarily traces cooler chromospheric and filament material. The 171~\AA{} channel captures plasma around $\sim$0.6 MK, highlighting quiet coronal loops. The 193~\AA{} channel is sensitive to both $\sim$1.25 MK for quiet corona and coronal holes, and $\sim$20~MK during strong flares. Sunspot and magnetogram data: We used vector magnetograms of NOAA Active Region 13947 acquired by the Helioseismic and Magnetic Imager (HMI) instrument onboard the Solar Dynamics Observatory (SDO) \cite{Schou2012}.
\subsection{Data Sources}
Different types of studies require different data sources. Several space and ground-based missions are currently observing the Sun. Data obtained from these diverse platforms enable us to study the Sun across multiple wavelengths, providing a comprehensive view from its surface (photosphere) to the outer layers of its atmosphere (chromosphere and corona).
\begin{itemize}
\item \textbf{GOES satellite (Geostationary Operational Environmental Satellite)}: The GOES satellite has a tool called XRS (X-ray sensor) that measures X-ray light from the Sun\cite{GOESXRS} . We used these measurements to see how strong and how long solar flares lasted. We basically used X-ray flux graph to elevate the research quality, and briefly discussed in Sect.~[\ref{sect3}].
    \item \textbf{Solar Dynamics observatory (SDO)}: We used data from SDO to study solar flare in different wavelengths. The following details are give below \cite{Pesnell2012}:
    \begin{itemize}
        \item \textbf{SDO AIA (Atmospheric Imaging Assembly)}: Extreme ultraviolet EUV images in 131\AA\, 171\AA\, 191\AA\, and 304\AA\. We used pictures from these specific light types:
        \begin{itemize}
            \item 131\AA\: This wavelength shows hot plasma in flares, particles and the hottest region.
             \item 171\AA\: This wavelength captures a quiet coronal region and magnetic loop.
             \item 193\AA\: This wavelength observes the corona and hot flare plasma.
             \item 304\AA\: This wavelength shows chromosphere and transition regions with strong emission from He{(+)} ions.
        \end{itemize}
       \item \textbf{SDO/HMI (Helioseismic and Magnetic Imager)}: HMI line-of-sight magnetograms and continuum intensity images are used to study sunspot evolution and magnetic polarity.

    \end{itemize}
    \item \textbf{SOHO MISSION (Solar and heliospheric observatory)} : The SOHO mission has a camera called LASCO(Large angle and spectrometric coronagrph) \cite{Domingo1995}. This camera watched for big explosions from the Sun called Coronal mass ejections (CMEs). We used the LASCO catalog for CME tracking. The catalog enables detailed scientific investigations of CMEs and their associated phenomena, such as solar flares.
    \item\textbf{The e-Callisto network}: The e-Callisto network data were used to identify and analyze solar radio bursts associated with flares and coronal mass ejections. The dynamic radio spectra provide information on the timing and evolution of energetic electron populations and shock-related phenomena in the solar corona, enabling comparison with signatures observed at other wavelengths \cite{Benz2009CALLISTO}.
    \item\textbf{Near-ultraviolet observations from SUIT/Aditya-L1}: Near-ultraviolet observations from SUIT/Aditya-L1 were employed to study the response of the lower solar atmosphere during eruptive solar events. Variations in NUV intensity and morphology were examined to investigate flare-related brightenings and active region evolution linked to CME initiation \cite{Ghosh2022SUIT}.
    \item\textbf{Hard X-ray data from STIX/Solar Orbiter}: Hard X-ray data from STIX/Solar Orbiter were used to investigate non-thermal electron acceleration during solar flares. Spectral and imaging information from STIX provides constraints on flare energetics and facilitates correlation with radio and NUV signatures associated with CME-driven activity \cite{Krucker2020STIX}.
    
\end{itemize}

\subsection{Tools Used}
All data are downloaded from SDO, AIA websites and then processed using python language \texttt{Fido} client and then after the retrieved images of different wavelengths have analyzed, observed and concluded by the team \cite{Lemen2012, SunPy2020}.

\subsubsection{GOES flux graph} GOES x-ray flux graph helped us to understand the changes of flux intensity with time and helped us to figure out the peak of the intensity at 12:47UTC  \cite{{Evans2010}}.
Custom scripts were developed for active region identification, Time-Series plotting, and AIA wavelength-based image analysis \cite{Garcia1994}.

\subsubsection{e-Callisto Network – Solar Radio Burst Detection} The e-Callisto network is used to monitor solar radio bursts that often accompany flares and CMEs. Radio signatures of the flare were examined using data from the e-Callisto network \cite{Benz2009CALLISTO}. These bursts provide early signatures of energetic electrons and shock formation in the solar corona during eruptive activity.

Analysis of dynamic radio spectra from e-Callisto allows the study of frequency drift patterns, which are closely related to electron beam propagation in flare events and shock speeds associated with CME-driven disturbances.

The wide geographical distribution of e-Callisto stations ensures continuous temporal coverage, making it possible to accurately relate radio burst onset times with flare emission and the early stages of CME evolution.

\subsubsection{Aditya-L1 – SUIT for Near-Ultraviolet Observations}: SUIT observations in the near-ultraviolet wavelength range are used to investigate how the lower solar atmosphere responds to flare and CME-related energy release. NUV emission reflects rapid radiative changes in the photosphere and lower chromosphere during solar activity. Near-ultraviolet observations were obtained from the SUIT instrument onboard Aditya-L1 \cite{Ghosh2022SUIT}.

Full-disk NUV images from SUIT enable detailed examination of flare-related brightenings, active region evolution, and footpoint emissions associated with erupting magnetic structures.

Continuous observations from the Sun–Earth L1 point allow SUIT to capture pre-flare, impulsive, and gradual phases of solar events, supporting reliable temporal correlation with coronal and high-energy signatures.

\subsubsection{Solar Orbiter – STIX Instrument for Hard X-Ray Observations} Hard X-ray data from STIX are used to study non-thermal electron acceleration during solar flares, a key process involved in flare energy release and CME initiation.

Spectral analysis of STIX observations provides information on electron energy distributions and the temporal evolution of flare energetics, offering insight into the efficiency of magnetic reconnection during eruptive events.

Imaging capabilities of STIX allow localization of hard X-ray sources in the corona, helping to connect flare acceleration sites with radio burst sources and CME trajectories. Hard X-ray emission was analyzed using measurements from the STIX instrument onboard Solar Orbiter \citep{Krucker2020STIX}.

\section{Results and Analysis}\label{sect3}
\subsection{Flare Energetics and Multi-Wavelength Light Curves}
\subsubsection{\textbf{GOES X-ray Observations}}
\begin{figure}[H]
    \centering
    \includegraphics[width=0.8\textwidth]{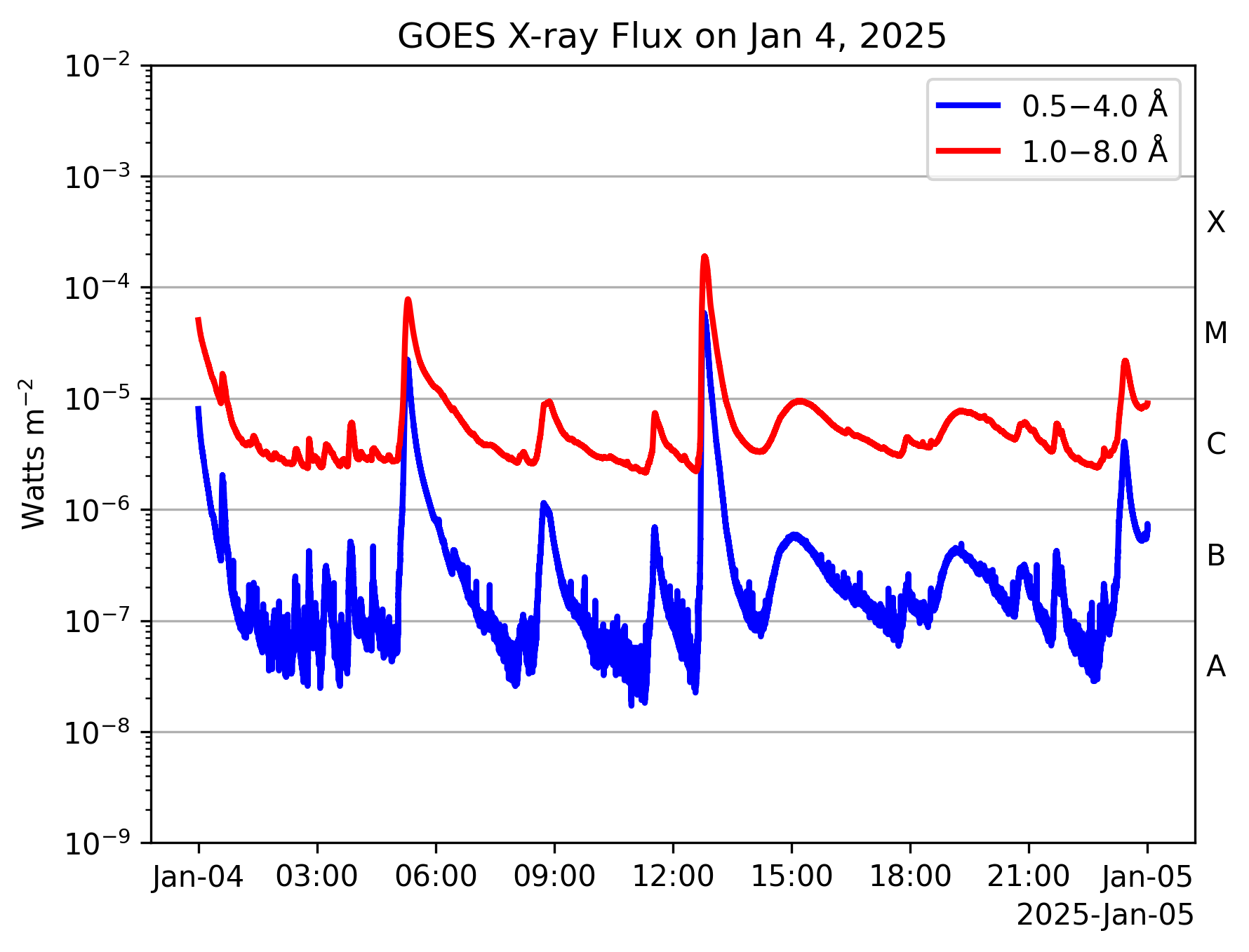}
    \caption{GOES X-ray flux  on 4 January 2025, show the two channels:- 1–8$Å$ (red) and 0.5–4.0\AA\ (blue). A peak around 12:47 UTC shows the detection of X1.8 class solar flare.}
    \label{fig1}
\end{figure}

The X-ray flux Fig.~(\ref{fig1}) show a rapid increase at 12:35 UTC, peaking at 12:47 UTC, confirming an X1.8-class flare.
This graph shows the X-ray light from the Sun, measured by GOES satellites on January 4, 2025. The blue line tracks X-rays in the 0.5–4.0 \AA range, and the red line tracks X-rays in the 1.0–8.0 \AA range. The scale on the right tells us how strong the flares are (A, B, C, M, X class) \cite{Garcia1994,White2005}.
Looking at the graph, we can see several peaks in solar activity throughout the day. The most significant event occurs around 12:47 UTC. At this time, the red line (1.0–8.0 \AA) sharply spikes up, reaching well into the "X" class region. This shows a very strong solar flare. Before this big flare, there's another noticeable peak just before 06:00 UTC, where the red line briefly enters the "M" class range.
After the major X-class flare around 12:47 UTC, the X-ray levels drop but remain fairly active. We see several smaller bursts where the red line climbs into the "M" class again, for example, around 14:00 UTC, and another notable one just before 24:00 UTC (midnight). Throughout the day, even when there aren't large flares, the red line generally stays in the "C" class range, indicating continuous, lower-level activity.
The blue line (0.5–4.0 \AA) generally follows the same pattern as the red line, but at much lower intensity. It shows similar peaks, like the big one around 12:47 UTC, but its values are always one or two classes lower on the scale. For instance, during the major X-class flare in the red line, the blue line reaches into the "M" class.
In short, the graph clearly shows (cf. Fig.~(\ref{fig1})) a very powerful X-class solar flare occurring around 12:47 UTC on January 4, 2025, alongside a series of other M-class and C-class flares that point to a highly active day on the Sun\cite{Lemen2012}. The total X-ray energy \(E_X\) of the X1.8-class solar flare on 2025 January 4 is estimated using:
\begin{equation}
  E_X = 4\pi D^2 \int_{t_1}^{t_2} F(t)\,dt ,  
\end{equation}
here the parameters are \(D = 1.496 \times 10^{13}\,\mathrm{cm}\), duration \(t_2 - t_1 = 1200\,\mathrm{s}\), average flux $(F(t) \approx 9 \times 10^{-2}\,\mathrm{erg\,cm^{-2}\,s^{-1}}$, we get
$E_X \approx 4\pi D^2 \times (F \cdot \Delta t) \approx 3 \times 10^{29}$ erg.

\subsubsection{\textbf{Thermal Evolution (AIA)}}

\paragraph{\textbf{Multi-Wavelength Light Curve}}

\begin{figure}[htbp]
    \begin{center}
     \includegraphics[width=0.85\textwidth]{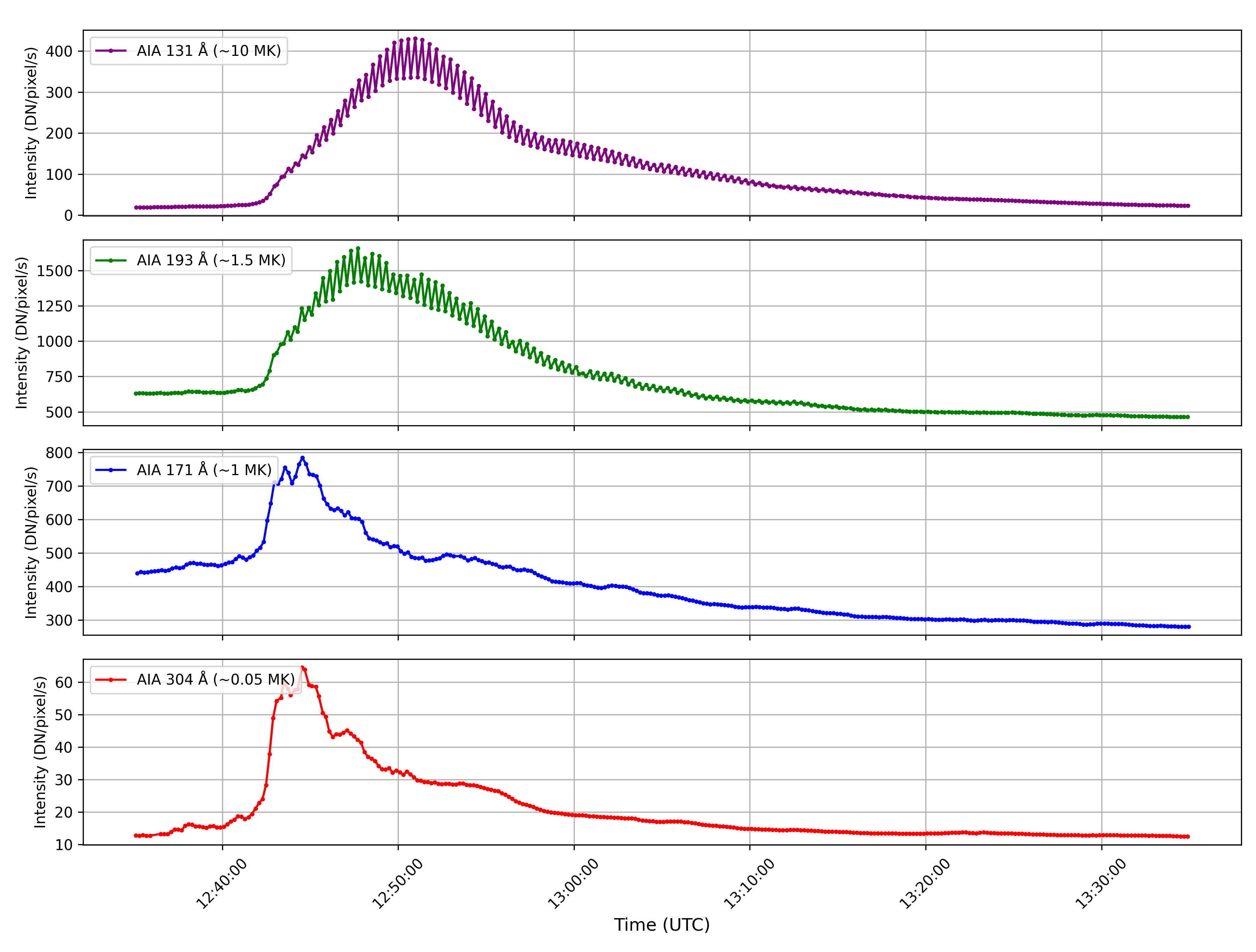}
    \end{center}
    \caption{Multi-thermal EUV light curves observed by SDO/AIA of the X.18 flare on January 4, 2025. Integrated intensities from 131 \AA\ ($\sim$10~MK),171 \AA\ ($\sim$1~MK), 193 \AA\ ($\sim$1.5~MK), \& 304 \AA\ ($\sim$0.05~MK) channels are displayed, presenting the coupled time response of the chromosphere \& corona. The 131 \AA\ hot channel peaks around the flare maximum, whereas the cooler choromospheric \& coronal channels shows delayed \& more gradual evolution, corresponding plasma heating \& cooling within the post-flare loop.}
    \label{fig:multiwavelength_lc}
\end{figure}

To understand the multi-thermal evolution of X1.8 flare, we plotted and analyzed the intensity profiles from four SDO/AIA channels - 131~\text{\AA}, 171~\text{\AA}, 192~\text{\AA}, 304~\text{\AA} (See Fig.~\ref{fig:multiwavelength_lc})\cite{Lemen2012}.
The light curves shows a close synchronous impulsive phase beginning around 12:42 UTC, followed by a peak at 12:47 UTC, in accordance with the GOES soft X-ray maximum \cite{Fletcher2011}.
In the intensity profiles, AIA 131~\text{\AA}
 is the hottest channel which is sensitive to ~10MK flare plasma, displays the most rapid intensity increase, tracking the super-heated plasma linked with magnetic reconnection \cite{Cargill1995}.
The chormospheric AIA 304~\text{\AA}
 channel shows the early, sharp intensification, indicative of a rapid lower atmospheric response, probably linked with the flare ribbon formation \& energy deposition by accelerated particles.
On the other hand, the cooler coronal channels - 171~\text{\AA}
 (~1.5 MK) \& 193~\text{\AA}
 (~1 MK), reveals a more gradual rise \& a slow decay phase.
Also, in 171~\text{\AA}
 emission the delayed decay is relative to the 131~\text{\AA}
 channel is consistent with the plasma cooling through radiative \& conductive processes.
The merged response across chromosphere, corona, \& the high temperature channels shows energy release in AR 13947 generated a coupled, multi-thermal response extending from the lower atmosphere to the hot corona.
\paragraph{\textbf{AIA 171 \AA\: Mapping the Structure of Magnetic Corona}}\label{sect3b1}
\begin{figure}
    \centering
    \begin{subfigure}[b]{0.8\textwidth}
        \centering
        \includegraphics[width=\textwidth]{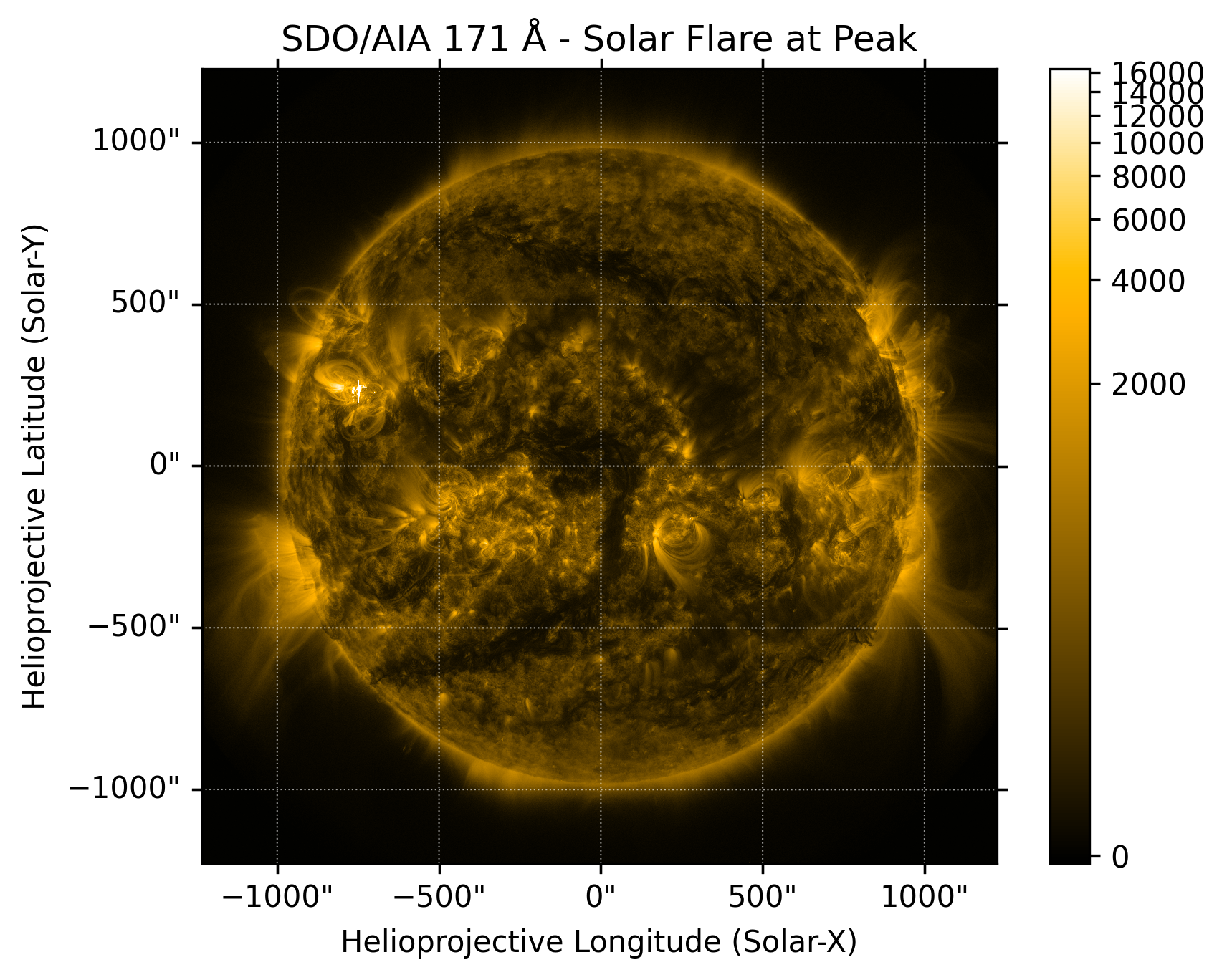}
        \caption{}
        \label{fig2a}
    \end{subfigure}
   \vspace{0.3cm}
    \begin{subfigure}[b]{0.49\textwidth}
        \centering
        \includegraphics[width=\textwidth]{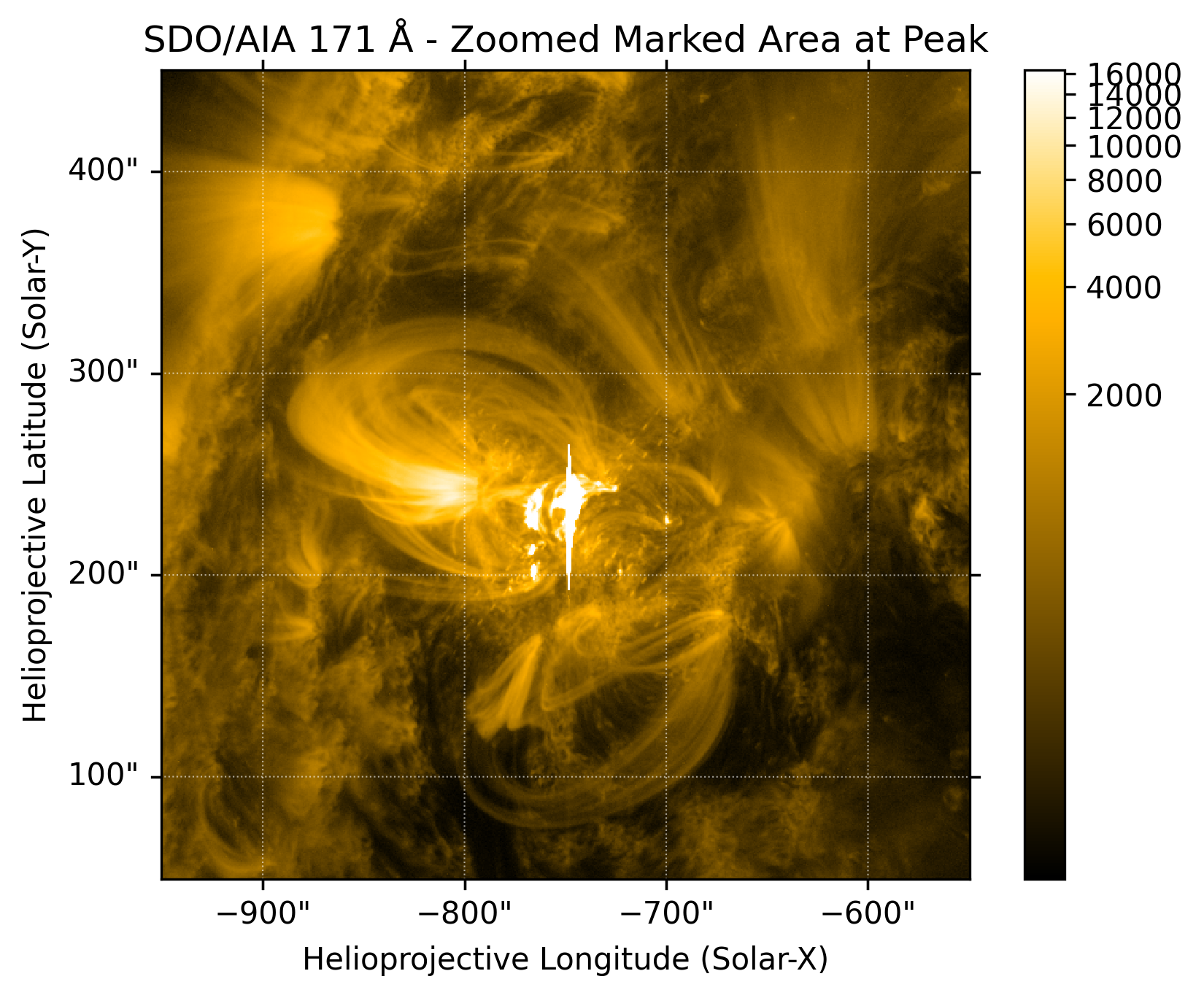}
        \caption{}
        \label{fig2b}
    \end{subfigure}
    \hfill
    \begin{subfigure}[b]{0.49\textwidth}
        \centering
        \includegraphics[width=\textwidth]{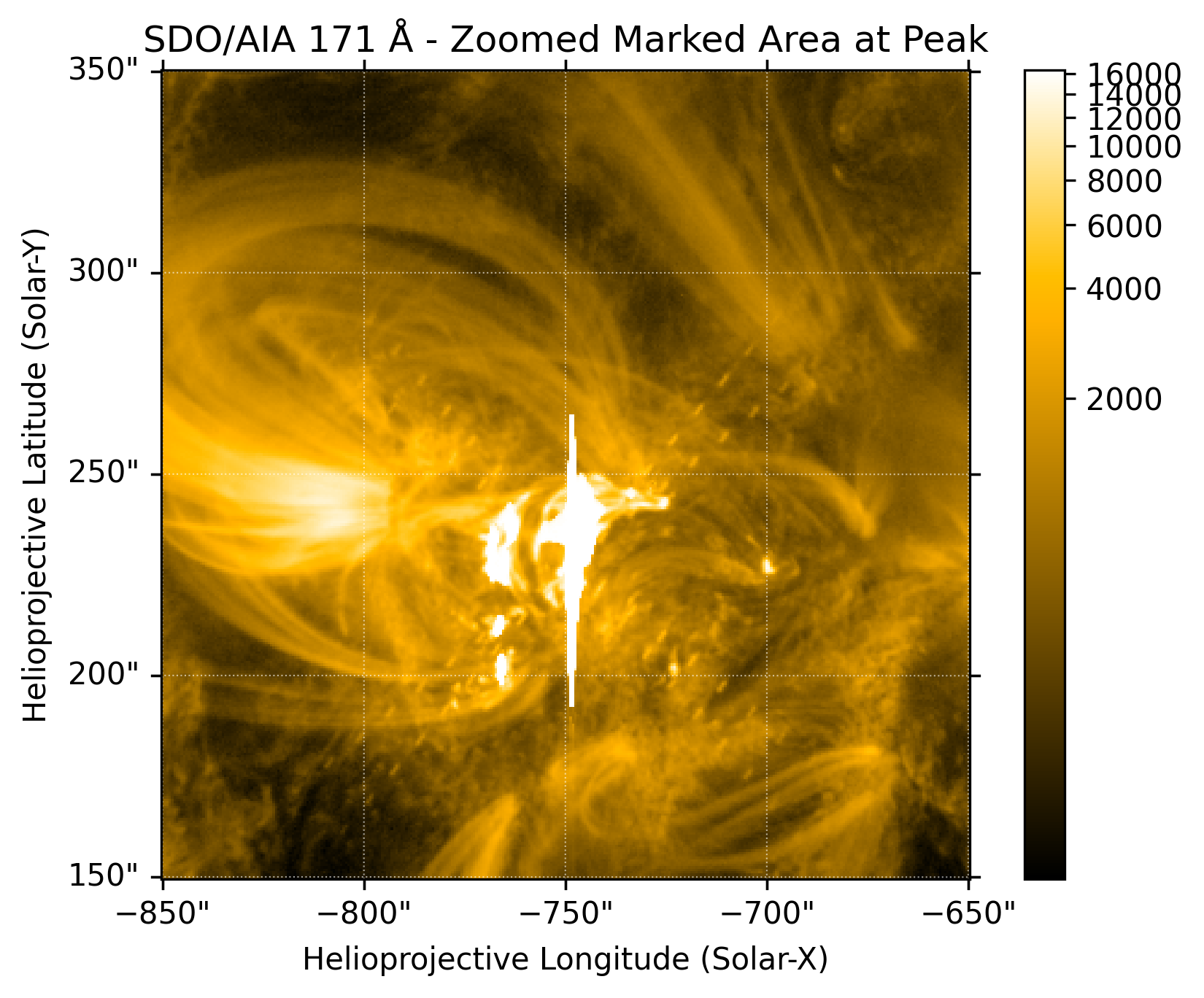}
        \caption{}
        \label{fig2c}
    \end{subfigure}
    \caption{(a)Full disk image observed by SDO/AIA 171 \AA\ at the X1.8 flare peak time on January 4, 2025, governed by emission from the coronal plasma at $\sim$1~MK temperature.(b) Magnified view of the erupting active region (AR13947), revealing the enhanced emission from coronal loop surrounding the flare erupting site. (c) Additional magnified view showing the fine scale morphology of coronal loops \& nearby plasma structure at flare peak.}
    \label{fig2}
\end{figure}
The 171 $Å$ band is mainly focused to track the emission lines of Fe IX, around ~0.6 million kelvin temperature it becomes more sensitive to the plasma. For mapping the pre-existing magnetic loop structures and its large-scale reformation following a flare, this cool coronal channel becomes more crucial \cite{DelZanna2011}.
\subparagraph{\textbf{Morphology of Pre-flare and Post-flare}}: 
Fig.~(\ref{fig2}) presents a full-disk image of the Sun. A prominent eruptive event - an X1.8-class solar flare is clearly visible in the north-western quadrant of the solar disk, centered at helioprojective coordinates Solar-X = $-600^{\prime\prime}$ and Solar-Y = $+300^{\prime\prime}$. The flare reached its peak intensity at approximately 12:47~UTC, with the overall event duration spanning from 12:35~UTC to 12:55~UTC. The erupted region is the Active Region (AR) 13947 exhibits an intense bright region which was caused by the magnetic reconnection process, further releasing the magnetic energy stored in the corona of sun appearing as $X$-class flare. However, the true advantage of this channel is the observation of the surrounding as a non-flaring plasma. It also visualizes a complex and dense network of the coronal loops sweeping across the whole solar disk that further connects to the different active regions as well as the quiet Sun area during the flare peak timing. Following the zoomed images as shown in  Fig.~(\ref{fig2b}) and  Fig.~(\ref{fig2c}) which is primarily dedicated to the AR 13947. These images helps us to observe and analyze the solar flare in a more comprehensive way, it shows the dense \& sophisticated system of interconnected magnetic loops around the erupted solar flare and the flare can be seen as a bright flash.
 In Fig.(\ref{fig2b}), the flare is located around Solar-X $\approx$
 -770” to -880” \& the Solar-Y $\approx$ 150” to 400”, sited at western limb at the Sun and, we  observe the flare further zoomed in as a flash appearing and located around approximately Solar-X $\approx$ -650” to -850” and Solar-Y $\approx$ 150” to 350” in  (cf. Fig.(\ref{fig2c})).
These two zoomed images help us to visualize the complexity and the direct manifestation of magnetic coronal loops sophistication which exhibits a region with high magnetic stress and some significant stored (free) magnetic energy which was released.
\subparagraph{\textbf{Post-Reconnection Loop Network:}} Following the transient phase of a flare, 'post-flare loops' are often formed. After the reconnection event these loops are interpreted to be the result of the magnetic field lines adjusting into more stable and lower-energy state. These effects are more efficiently observed by the AIA 171~$\AA$ channel as they cool \cite{Forbes1996}. The arch-like structure visible in Fig.~(\ref{fig2b}) and Fig.~(\ref{fig2c}) is stable with a newly formed arcade of loops. Further, the clear evidence of large-scale magnetic reforming was explained by the loop frame, in strong contrast to the more chaotic magnetic arrangement expected before the flare event. To estimate the amount of plasma heated and infused into them during the flare event, we can further use the brightness and density of these loops.\\

\paragraph{\textbf{AIA 193~\AA: Monitoring the Coronal effect and Hot Flare Plasma}}
The emission of Fe XII around 1.25 million kelvin and Fe XXIV around 20 million kelvin dominates the 193 Å channel during the flares, acts as a pathway between the intensely heated flare plasma and quiescent corona \cite{ODwyer2010}.

\begin{figure}
    \centering
    \begin{subfigure}[b]{0.8\textwidth}
        \centering
        \includegraphics[width=\textwidth]{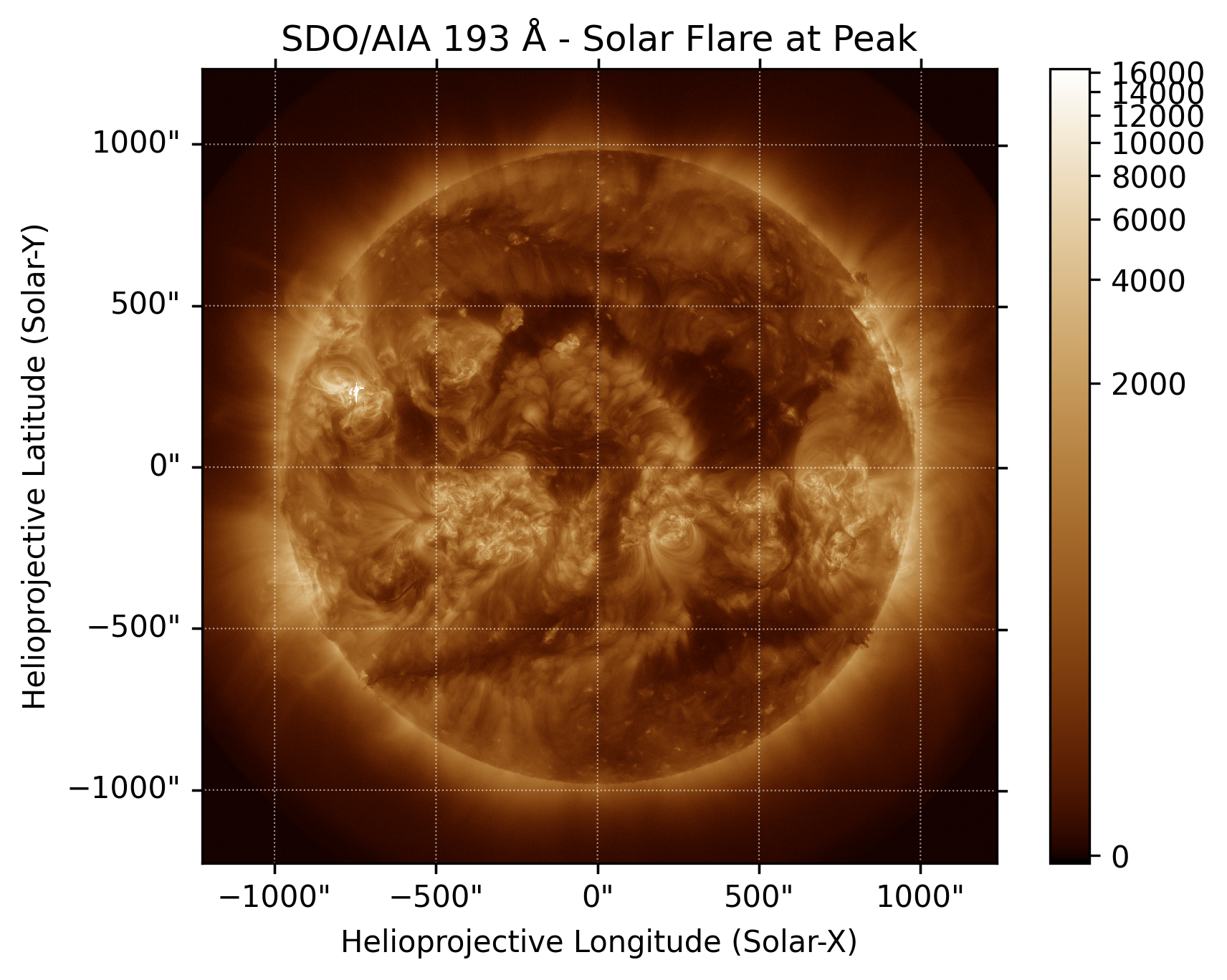}
        \caption{}
        \label{fig3a}
    \end{subfigure}
    \vspace{0.3cm}
    \begin{subfigure}[b]{0.49\textwidth}
        \centering
        \includegraphics[width=\textwidth]{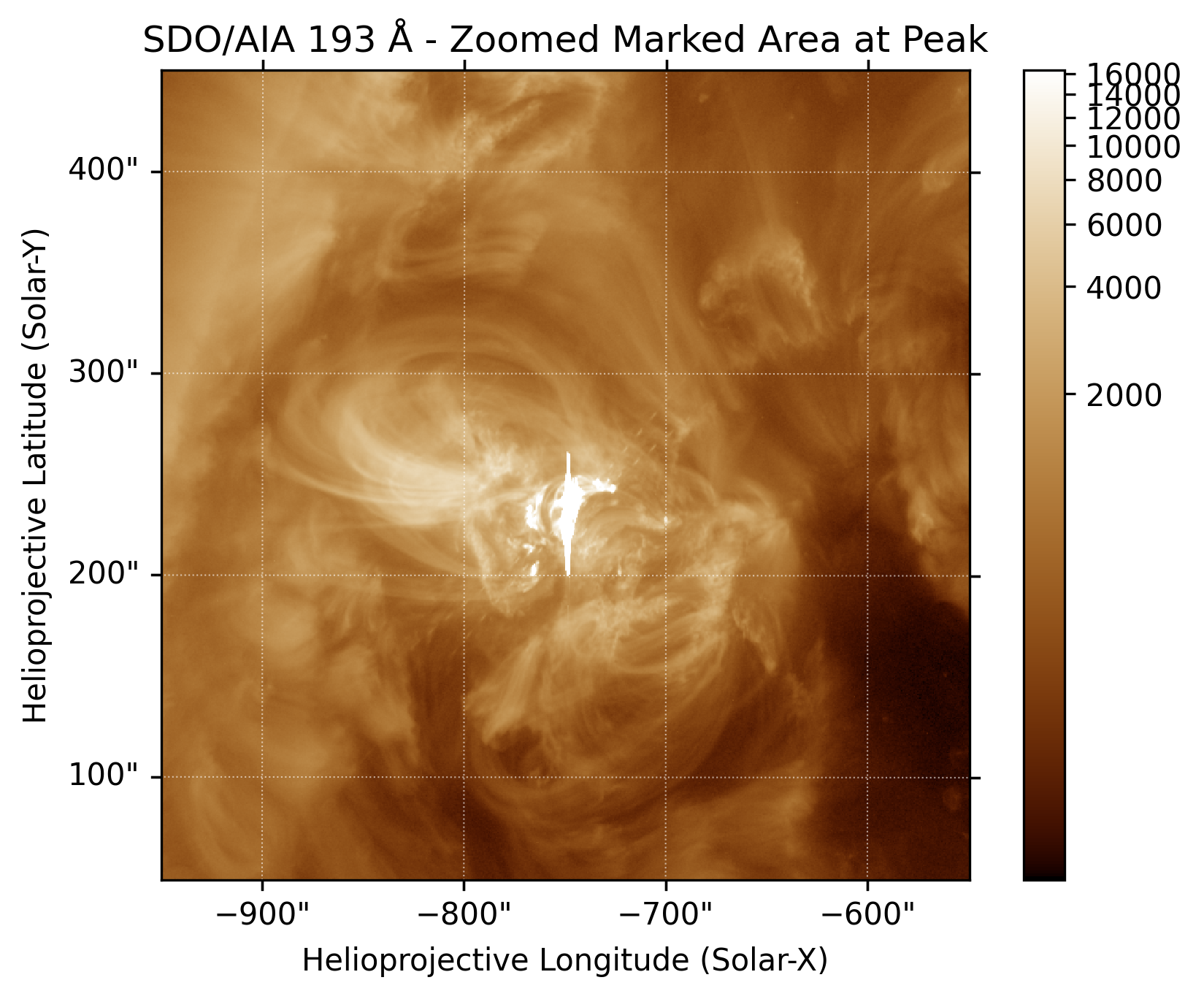}
        \caption{ }
        \label{fig3b}
    \end{subfigure}
    \hfill
    \begin{subfigure}[b]{0.49\textwidth}
        \centering
        \includegraphics[width=\textwidth]{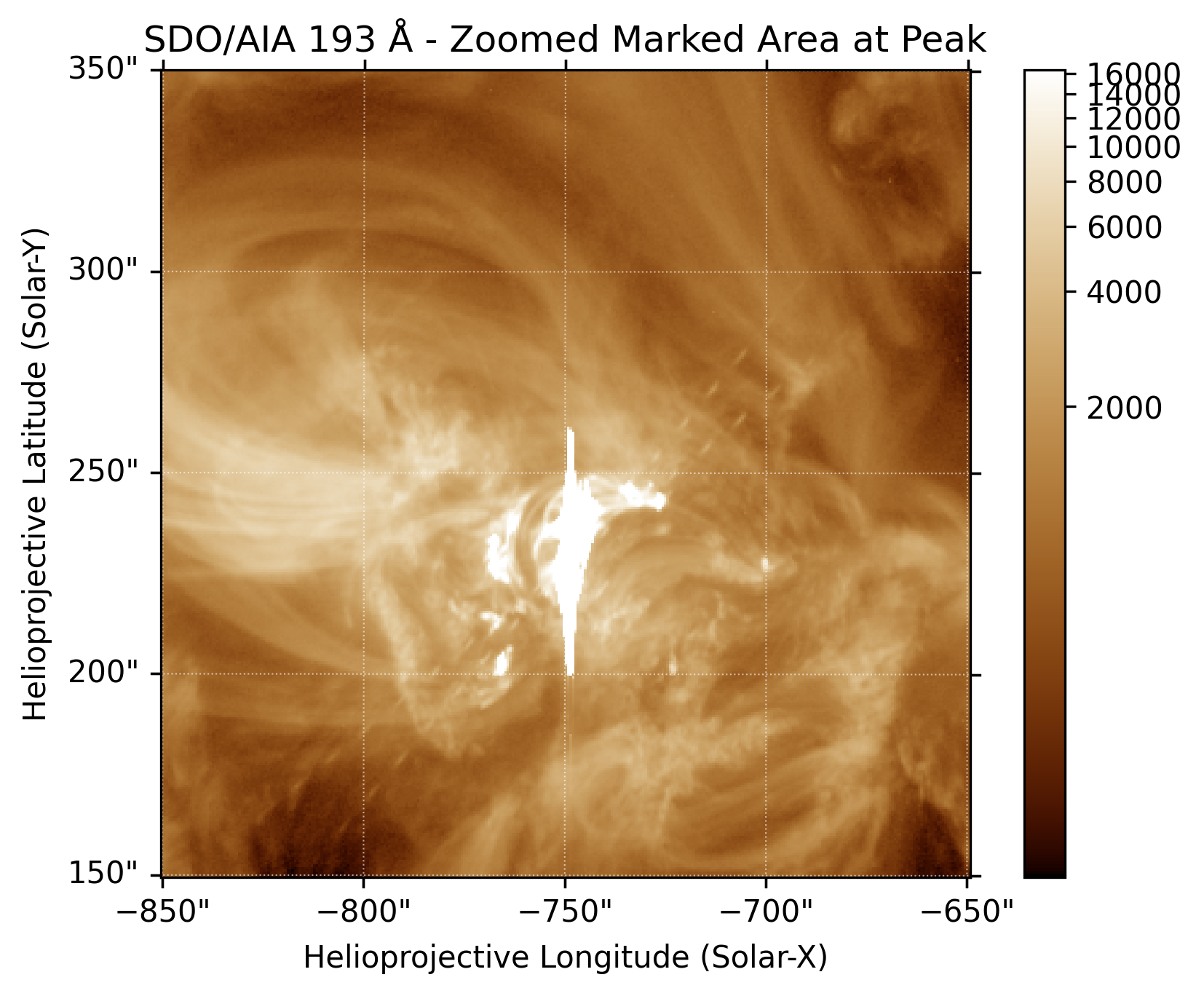}
        \caption{}
    \end{subfigure}  
    \caption{(a) Full disk image captured by SDO/AIA 193 \AA\ at the X1.8 flare peak time on January 4, 2025, mainly sensitive to coronal plasma at $\sim$1.5~MK temperature, with additional influence from hotter flare plasma. (b) Magnified view of the flaring Active region (AR13947), exhibiting the enhanced coronal emission \& arrangement of varying loops surrounding the flare location. (c) Additional magnified view indicating the complex morphology of coronal loops and nearby plasma structures linked with the flare.}
    \label{fig3c}
\end{figure}

\subparagraph{\textbf{Flare foot point  and Ejecta:}} As we can see in the Fig.~(\ref{fig3a}), the behavior of the flare in this channel shows its bright feature, dominant, marking the location of prominent heating of plasma at the same region of AR 13947 (north western region). Further in Fig.~(\ref{fig3b}) \& Fig.~(\ref{fig3c}), we can see the zoomed in region at the approximately same helio-projective coordinate as mentioned in Sect.~(\ref{sect3b1}), which shows the brightening and extended structures which indicates the region of the footpoints and the flare kernels where the energy is accumulated in the lower corona/transition region. The stretched, curved formations arching above the bright core shows the flare arcades formed by subsequent reconnection of magnetic field lines. This channel also effectively detected the bulk of heated coronal plasma.
\subparagraph{\textbf{Coronal Dimming:}} One of the most critical phenomena observable in this channel is the "coronal dimming”. This signifies to the reduced EUV localized region emissions that can be seen near a flaring active region. Dimmings are mostly investigated as the depletion of density occurred by the evacuation of plasma during the ejection of Corona mass ejection (CME)\cite{Hudson1996, Thompson2000}.\\

\paragraph{\textbf{AIA 304 \AA: The Chromospheric signatures of Energy Deposition}}
In AIA 304 \AA, the He II emission line dominates, it shows the plasma in the chromosphere and the transition region  at temperature around ~50,000 kelvin. Mostly, it highlights the lower atmospheric influence of coronal energy release \cite{Lemen2012}.
\begin{figure}
    \vspace{0.3cm}
    \begin{subfigure}[b]{0.49\textwidth}
        \centering
        \includegraphics[width=\textwidth]{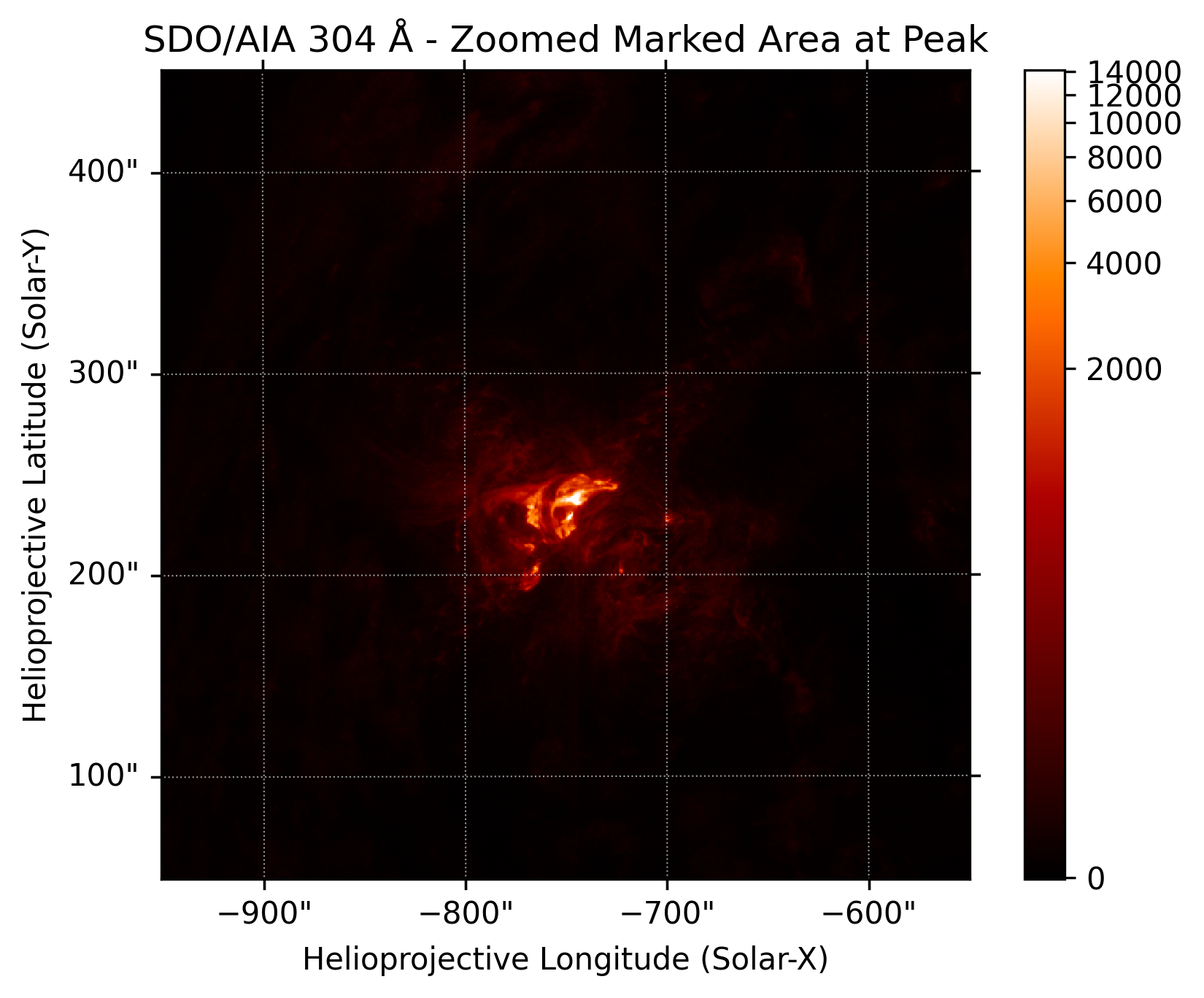}
        \caption{}
        \label{fig4b}
    \end{subfigure}
    \hfill
    \begin{subfigure}[b]{0.49\textwidth}
        \centering
        \includegraphics[width=\textwidth]{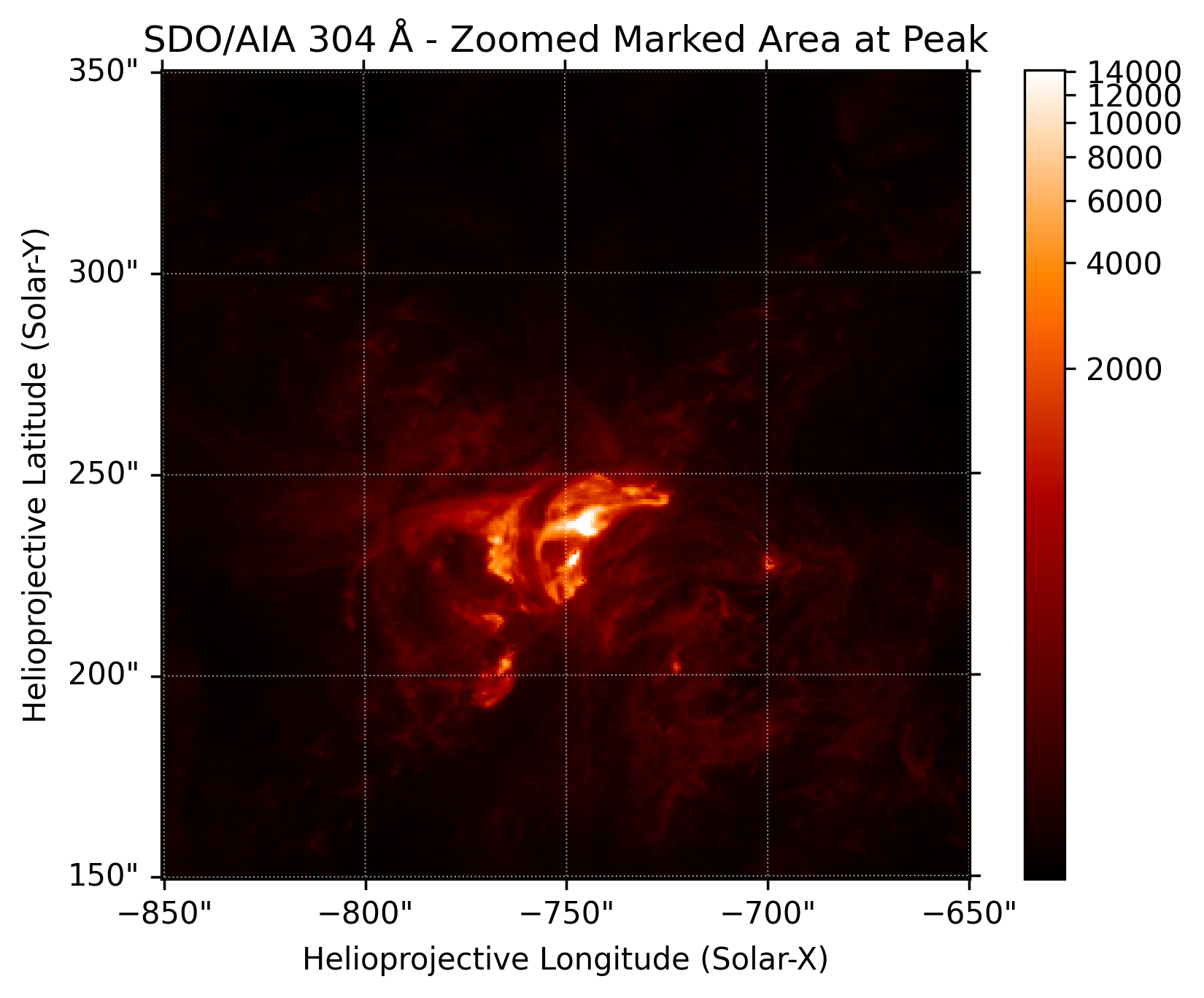}
        \caption{}
        \label{fig4c}
    \end{subfigure}
    \caption{(a) Magnified view of the erupting active region (AR13947), revealing the enhanced chromospheric emission linked with the flare. (b) Additional magnified view showing the fine-scale morphology of the near surface plasma structure \& ribbon-like emission emerged during the flare.}
    \label{fig4}
\end{figure}
\subparagraph{\textbf{Flare Ribbons:}} The most prominent features of the band 304 \AA , Fig.~(\ref{fig4}) shows two bright, prolonged structures known as the "flare ribbons." These are the chromospheric signatures of the newly merged post-flare loops. Throughout the process of magnetic reconnection in the corona, the high-energy particles which (protons and electrons) are accelerated. Further these particles transit down the magnetic field lines and encounters with the chromospheric denser plasma, which leads it to heat rapidly and radiate strongly, resulting the observed ribbons.
\subparagraph{\textbf{Morphology and Evolution:}} In Fig.~(\ref{fig4b}) and Fig.~(\ref{fig4c}) the morphology of the ribbon can be seen as a precise projection of the coronal magnetic loop geometry. By the inspecting the separating of the flare ribbons and their motion apart from each other throughout the flare’s evolution is a crucial prediction of the standard solar flare model \cite{Shibata2011}. It indicates the reconnection site shifting to higher altitude into the corona over time, incorporating the expanding number of magnetic field lines.\\

\paragraph{\textbf{AIA 131 \AA: Mapping the Super-Heated Flare Core}}

The 131~{\AA}   channel with primarily Fe VIII and Fe XXI is responsive to extremely hot plasma, with a major contribution from plasma above 10 million Kelvin during the eruption of a flare. This makes it a superior tool for the isolation of the hottest and most energetic parts of the occurrence \cite{ODwyer2010}.

\begin{figure}[H]
    \begin{subfigure}[b]{0.49\textwidth}
        \centering
        \includegraphics[width=\textwidth]{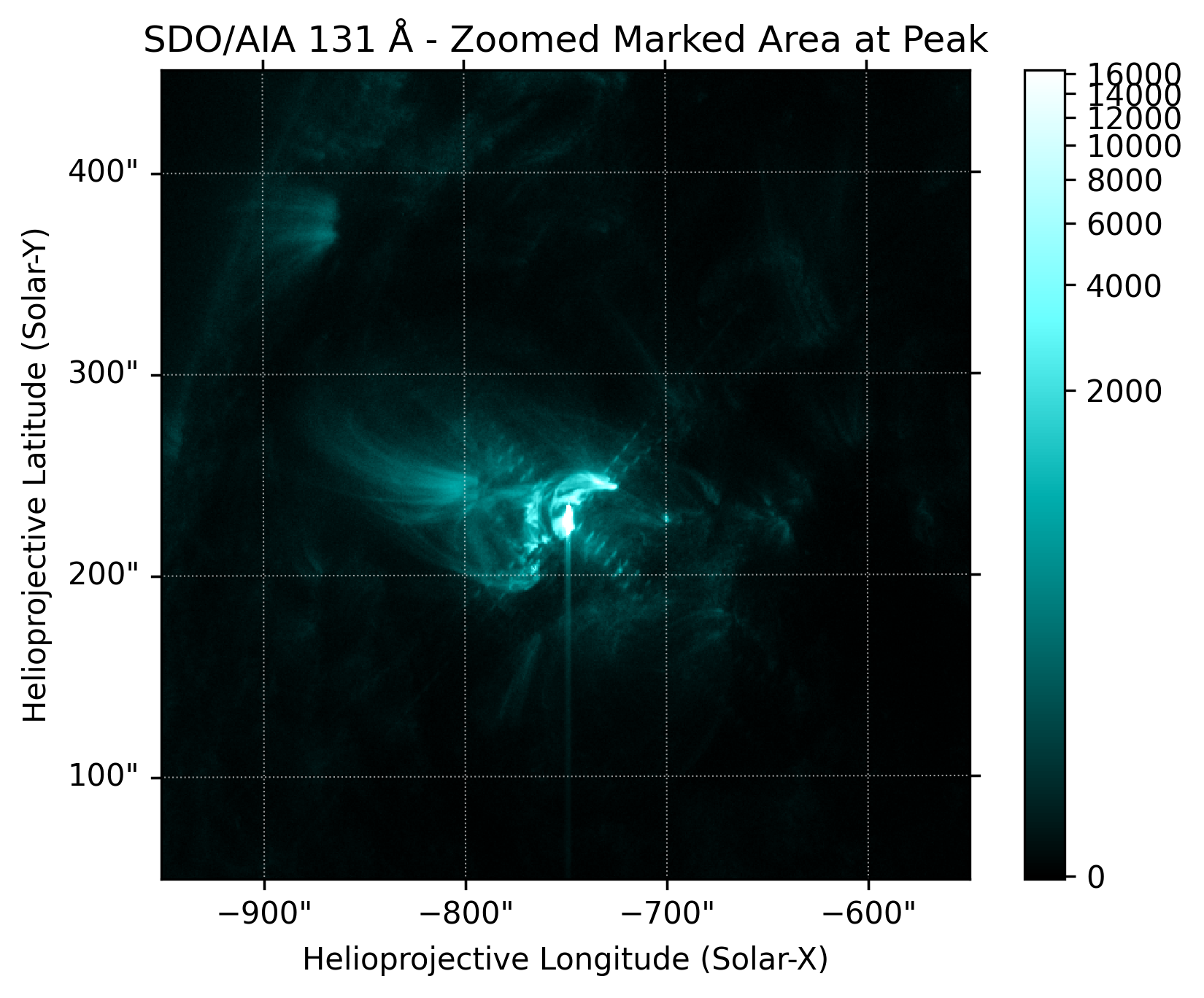}
        \caption{}
        \label{fig5b}
    \end{subfigure}
    \hfill
    \begin{subfigure}[b]{0.49\textwidth}
        \centering
        \includegraphics[width=\textwidth]{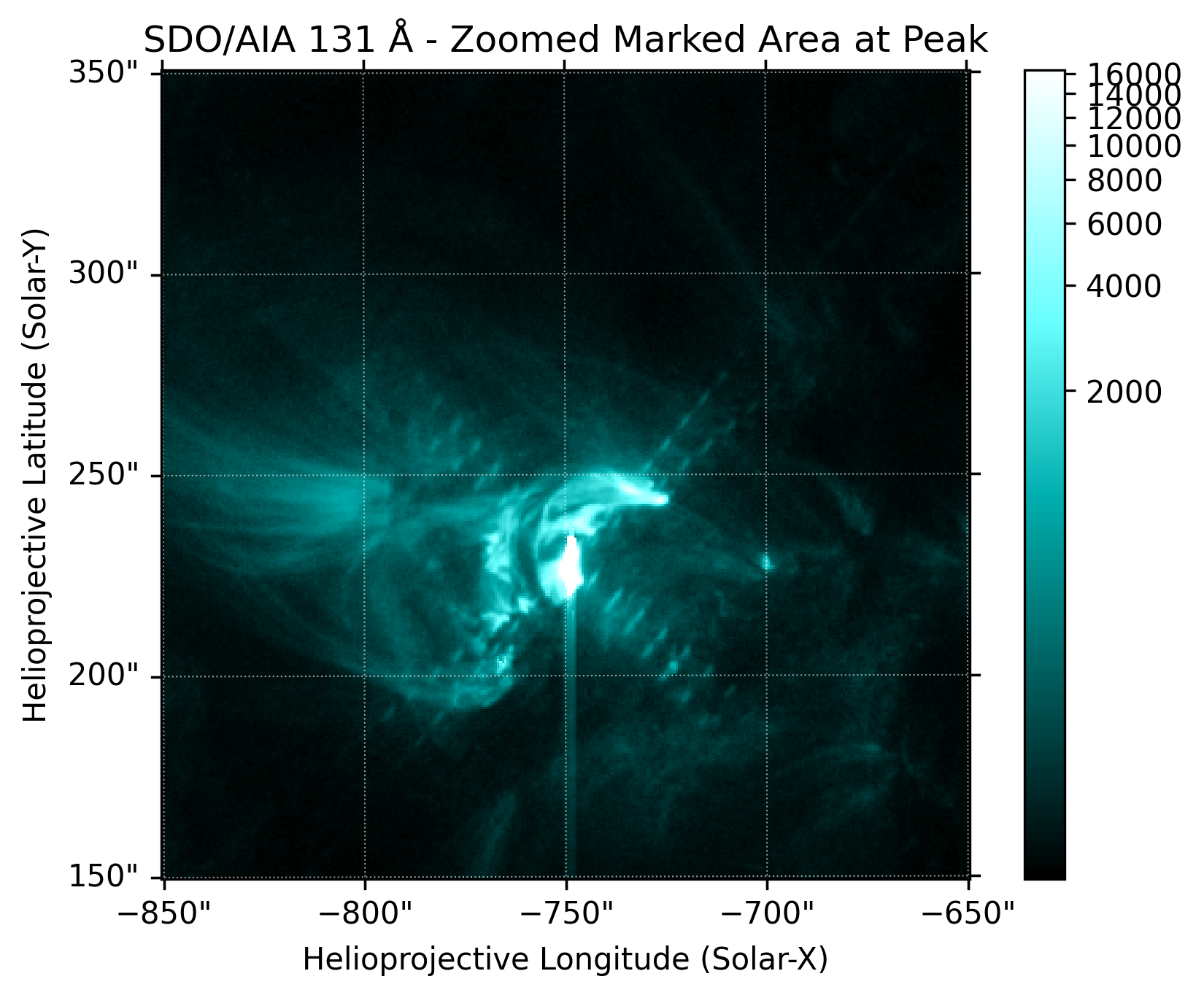}
        \caption{}
        \label{fig5c}
    \end{subfigure}
    
    \caption{(a) Magnified view of the erupting active region, displaying the enhanced coronal emission linked with the flare. (b) Additional magnified view of same active region, showing the fine-scale arrangement of hot coronal loops.}
    \label{fig5}
\end{figure}
\subparagraph{\textbf{Region of Magnetic Reconnection:}} In the Fig.~(\ref{fig5}), it shows the flare in the 131 {\AA} band. In contrast to the more diffuse configuration seen before at cooler temperatures, here the emission is highly concentrated and extreme. The bright central region In Fig.~(\ref{fig5b}) and Fig.~(\ref{fig5c}), is considered to mark the location of the primary energy discharge and the region of magnetic reconnection itself, or the super-heated peaks of flare loops. In these regions, the plasma is heated at very high temperatures ($>$10 million Kelvin) almost abruptly.

\subparagraph{\textbf{Flare Profile and Energetics:}} The fine morphology within this hot channel shows the sophisticated magnetic configuration at the flare’s core. Also several small scale flare loops and radiant knots can often be identified, corresponding to distinct magnetic reconnection events. The extreme and dense emission is aspect of a powerful $X$-class solar flare, which confirms that a notable amount of magnetic energy was converted into thermal and kinetic energy into very small space.
\\
\subsubsection{\textbf{Near-Ultraviolet Range (NUV)}}

\paragraph{ADITYA L1 - SUIT}

\begin{figure}[htbp]
    \centering
    \includegraphics[width=0.75\textwidth]{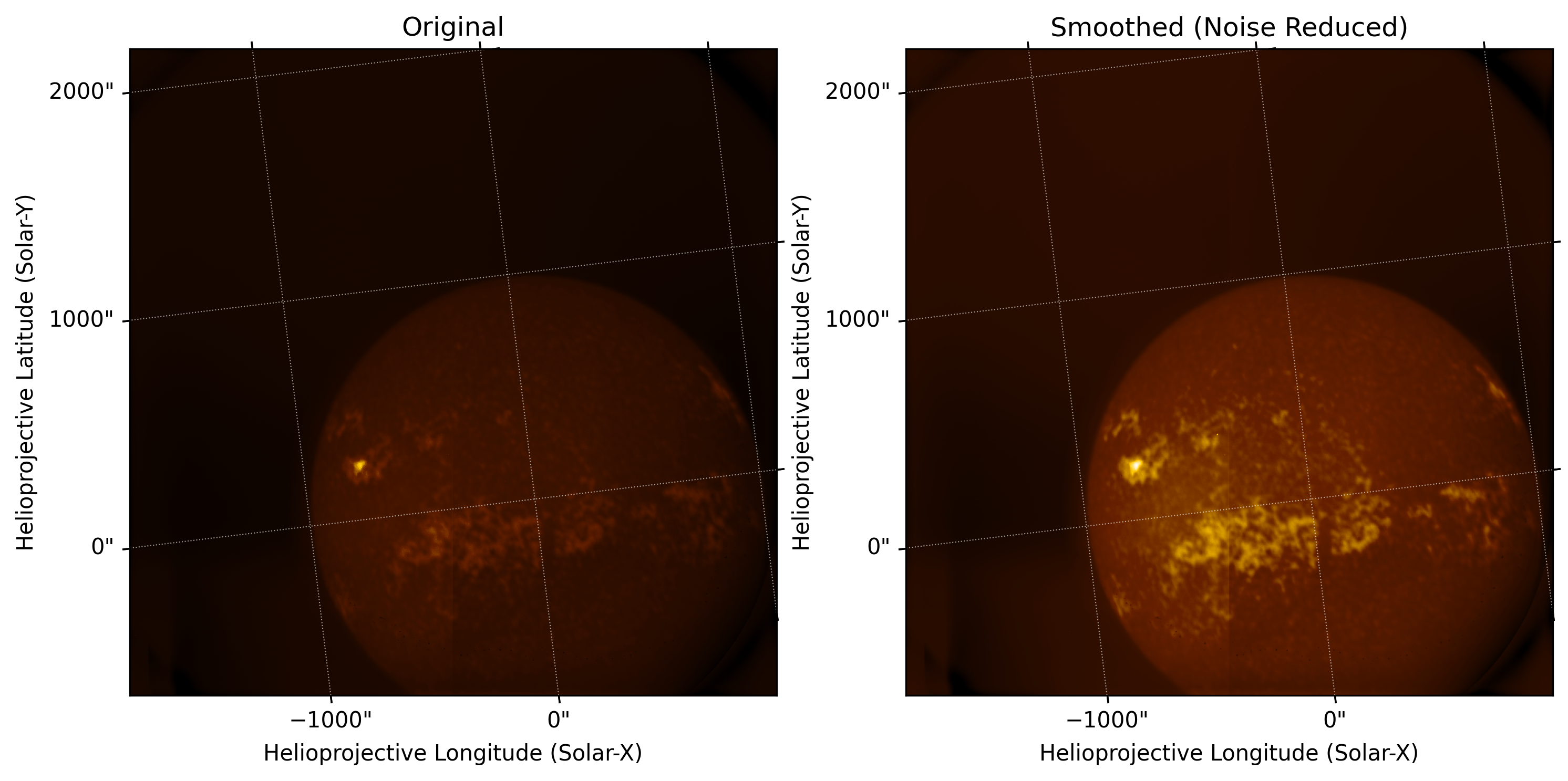}
    \caption{Near-Ultraviolet full-disk image (NUV) image of the Sun captured SUIT instrument onboard Aditya-L1 during the X1.8 flare on January 4, 2025. The left panel displays the original observation \& the right panel shows a smoothed version for enhanced better visualization. In the northwestern quadrant of Active Region 13947, enhanced NUV emission is visible.}
    \label{fig:aditya_l1_full_disk}
\end{figure}

To further augment our multi-wavelength analysis, we incorporated Near Ultra-Voilet observations (NUV) from SUIT instrument – onboard Aditya-L1 mission.
The transition region \& the corona is mostly focused by AIA channels, in addition SUIT provides the crucial insights into the lower atmosphere (upper photosphere-chromosphere) where the initial energy deposition of the solar flare occurs \cite{Hudson1996}. The full disk view of the Sun in Fig.~(\ref{fig:aditya_l1_full_disk}) was captured during the 4 January 2025 event. A noticeable area of intense UV brightening is visible in the north-western quadrant, which correspond to the AR 13947.
 We applied a noise reduction smoothing algorithm [see Fig.~(\ref{fig:aditya_l1_full_disk})] to improve the visual clarity of the active region.By comparing the original \& smoothed images, it reveals the intensified contrast behavior of the flaring region versus the quiet Sun background.In NUV range, the continuous brightening reveals sustained heating in lower atmosphere, confirming the X1.8 flare was a deep-seated eruptive event which coupled the lower atmosphere and the corona.

\subsection{The Flare Trigger I: Magnetic Flux Dynamics (HMI)}

\begin{figure}
        \centering
        \includegraphics[width=0.7\textwidth]{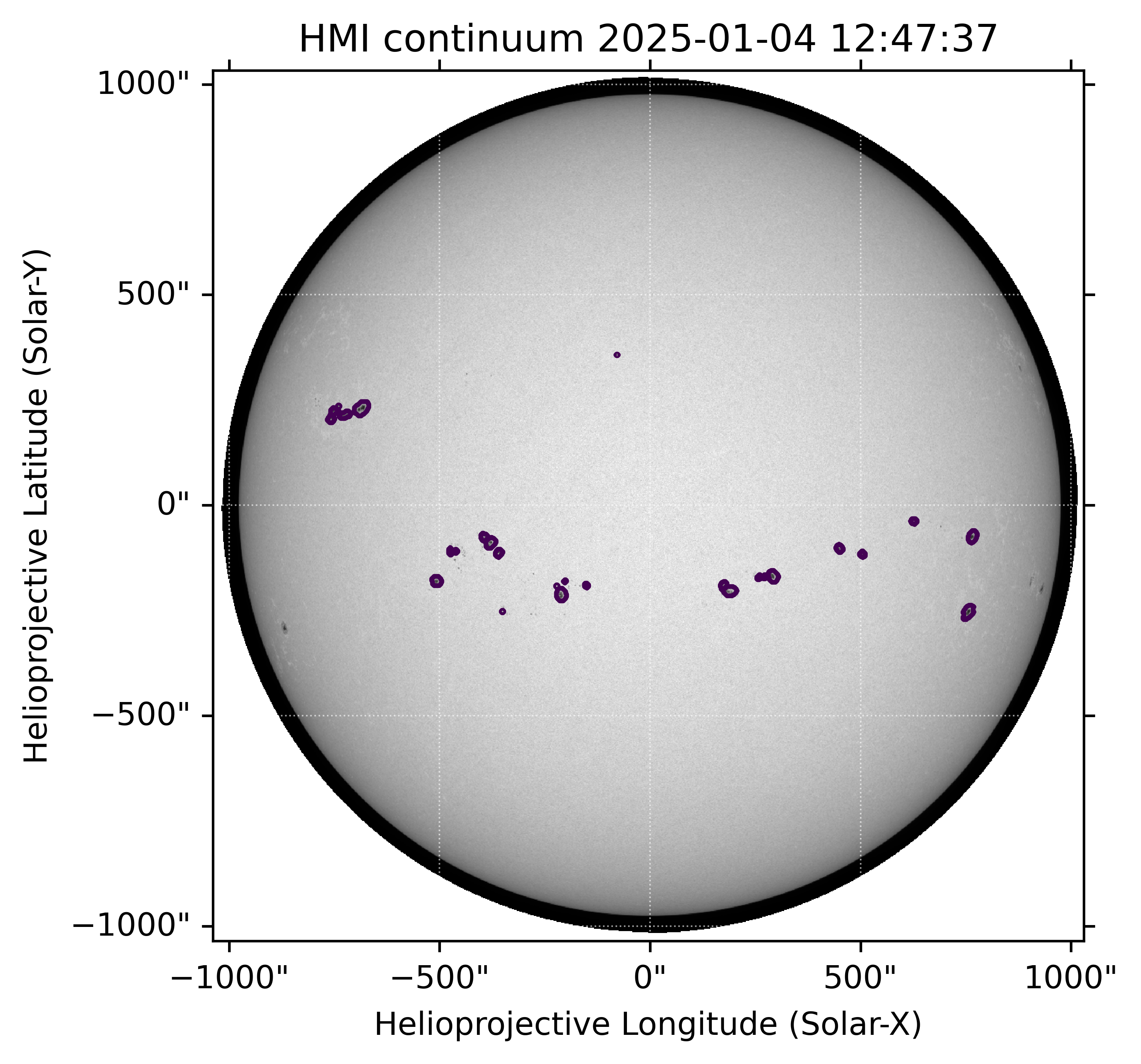}
        \caption{Displays the full-disk SDO/HMI continuum image at 12:47:37 UTC (peak flare time) on January 4, 2025, showing the detected sunspots outlined by purple contours.}
        \label{fig7}
\end{figure}

The Helioseismic and Magnetic Imager (HMI) onboard Solar Dynamic Observatory (SDO) enables the line-of-sight magnetic field and intensity continuum images and also allows precise monitoring of sunspot evolution and morphology of the solar photosphere, as the primary factor for the occurrence of solar is the rapid ejection of free magnetic energy which is stored in the solar corona \cite{Schou2012}. We performed a detail analysis of the magnetic behavior for AR 13947 which is the flare's source region.
Fig.~(\ref{fig7}) shows the full-disk HMI continuum image observed at 12:47:37 UTC, corresponding with flare’s peak, validates the active region 13947 situated in the northwestern quadrant was dominant and complex active region on the visible solar disk around that time.

\begin{figure}[H]
    \centering
    \begin{subfigure}[t]{0.48\textwidth}
        \centering
        \includegraphics[width=\linewidth]{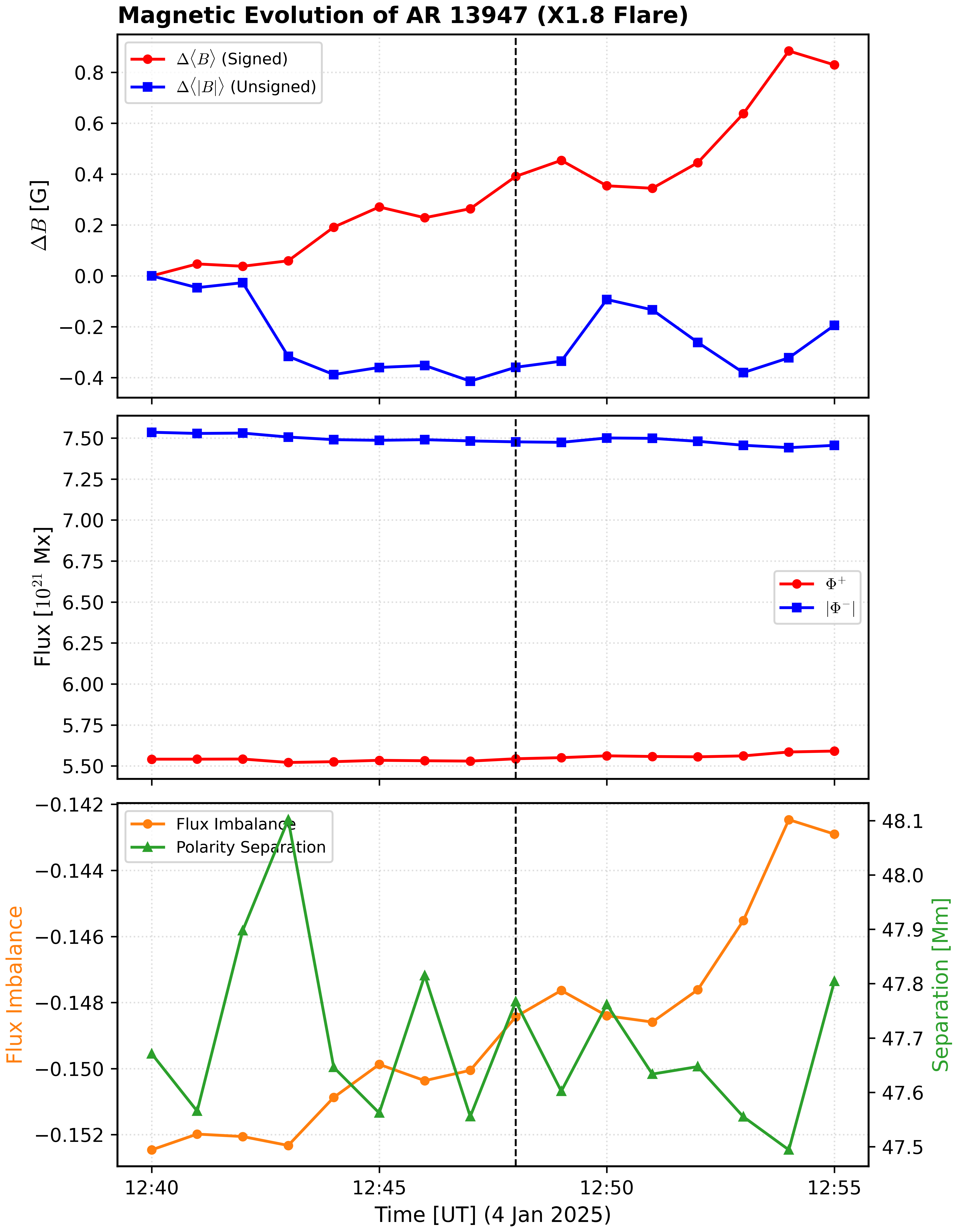}
        \caption{}
        \label{Evolution}
    \end{subfigure}
    \hfill
    \begin{subfigure}[t]{0.48\textwidth}
        \centering
        \includegraphics[width=\linewidth]{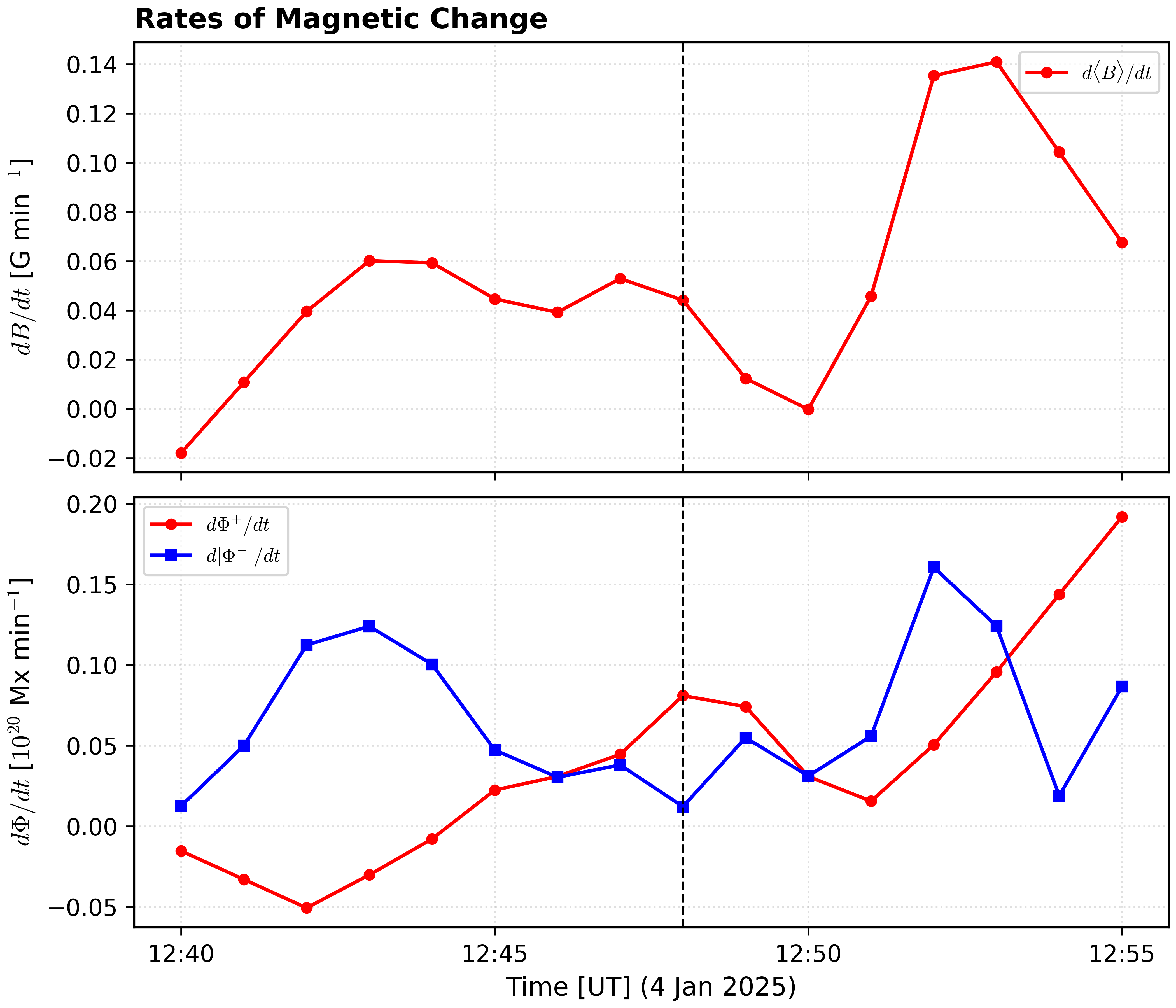}
        \caption{}
        \label{Derivative}
    \end{subfigure}

    \caption{(a)Shows the rates of photospheric magnetic variation around AR 13947 derived from SDO/HMI magnetograms, revealing the temporal evolution of mean magnetic field change rate $(d\langle B\rangle/dt)$ \& the signed and unsigned flux derivatives $(d\Phi^{+}/dt,\; d|\Phi^{-}|/dt)$. (b) The cumulative magnetic evolution, such as the change in signed \& unsigned magnetic field strength $(\Delta\langle B\rangle,\; \Delta|B|)$, total positive \& negative magnetic flux $(\Phi^{+},\; |\Phi^{-}|)$, \& the associated flux imbalance \& polarity separation. The peak flare time (12:48 UTC) is marked by the vertical dashed line over all panels, showing the abrupt and permanent magnetic restructuring linked with the eruption.}
\end{figure}

To identify the magnetic evolution linked with X1.8 flare, we examined the photospheric magnetic field measurements of AR 13947 using the SDO/HMI vector magnetogram (see Figure~\ref{Evolution} \& ~\ref{Derivative}).
The unsigned magnetic flux stayed nearly constant during the observational time range, revealing the absence of significant flux emergence or cancellation throughout the event \cite{Sudol2005}.
In contrast, the temporal evolution is exhibited by the signed magnetic flux, with the positive flux of $\sim \num{5.5e21}\,\text{Mx}$ compared to the dominating negative flux at $\num{7.5e21}\,\text{Mx}$ (see Figure~\ref{Evolution}, middle panel).
We also measured some substantial variations in the flux imbalance \& the polarity separation (see Figure~\ref{Evolution}, bottom panel).
Especially, the increased polarity separation near 12:43 UTC, several minutes before to the rapid phase, indicating enhanced magnetic shearing or stressing in the active region preceding the eruption \cite{Wang1994}.
The impulsive photospheric response of the flare is emphasized by the temporal derivatives of the magnetic parameters (see Figure~\ref{Derivative}) \cite{Schrijver2007}.
Whereas, the rate of change of mean magnetic field $\left(\frac{d\langle B\rangle}{dt}\right)$ stayed relatively modest prior the flare, it increased following the flare peak time at 12:48 UTC, further reaching a maximum of $0.14\,\mathrm{G\,min^{-1}}$ at 12:53 UTC.
Simultaneously, the strong post flare enhancement is presented by the signed magnetic flux derivative $\left(\frac{d\Phi^{+}}{dt}\right)$, while the unsigned flux derivative $\left(\frac{d\lvert \Phi\rvert}{dt}\right)$ remained comparatively constant, reinforcing the interpretation that the observed evolution reflects magnetic field rearrangement rather than the net flux change \cite{Petrie2012}.
Collectively, these changes in the flux imbalance, magnetic field, polarity separation are consistent with abrupt reconfiguration of the photospheric magnetic field following energy release, often regarded as the surface signature of coronal magnetic reorganizing \& a possible downward Lorentz-force impulse imposing on the lower atmosphere \cite{Hudson2008,Fisher2012}.

\subsection{The Flare Trigger II: Sunspot Morphological Evolution}

Its morphology helps us to identifies the dense grouping of multiple sunspots in the region of flare occurrence, which indicates the high magnetic complexity of delta-class, which is recognized to be highly flare-productive \cite{Zirin1987}.

\begin{figure}[H]
        \centering
        \includegraphics[width=0.75\textwidth]{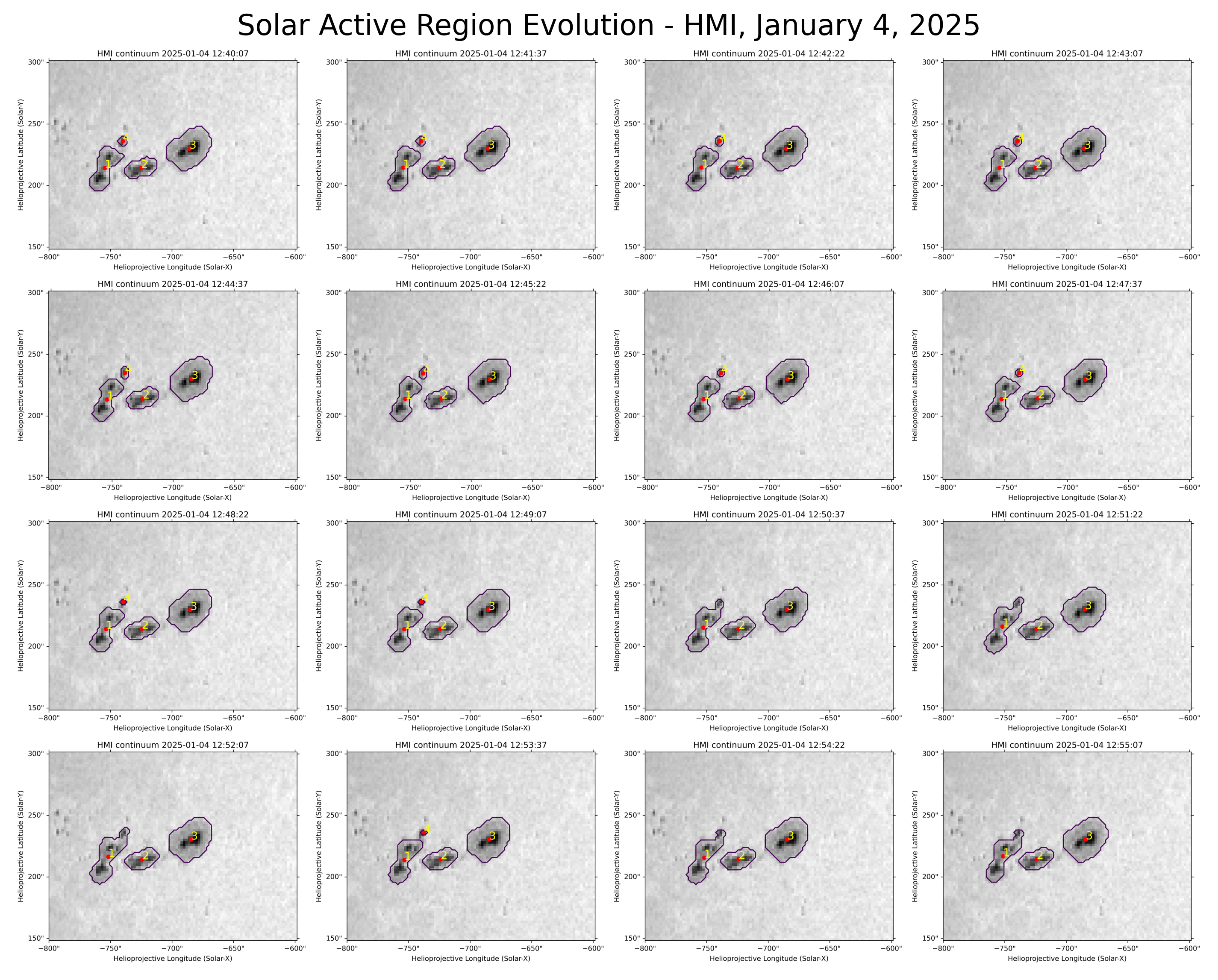}
        \caption{Shows the temporal sequence HMI continuum subframes revealing the morphological evolution changes of sunspot groups in Active Region 13947 throughout the pre-flare \& post-flare interval, with consistent purple contouring \& red centroid-used to track are \& the change in shape.}
    \label{solar_active_region_grid_highres}
\end{figure}

A more detailed view of the active region 13947 can be seen in Table~\ref{table:sunspot_data}, which is shown in time series of submaps ranging from 12:40 to 12:55 UTC. These images indicate the highly active environment, the sunspots are not stationary as they manifests subtle but notable changes in their positions and shape over the interval of 15-minutes which was the solar flare time range Fig.~(\ref{solar_active_region_grid_highres}). This suggests the existence of magnetic shearing and the rotational motions at the photospheric level, further this inject infuse magnetic stress and energy into the corona which eventually powers the flare. The quantitative measurements of the four sunspots enclosed in AR 13947 is presented in Table~(\ref{table:sunspot_data}) and illustrated in the following-used to track are \& the change in shape.

\begin{figure}[H]
    \centering

    \begin{subfigure}[t]{0.48\textwidth}
        \centering
        \includegraphics[width=\linewidth]{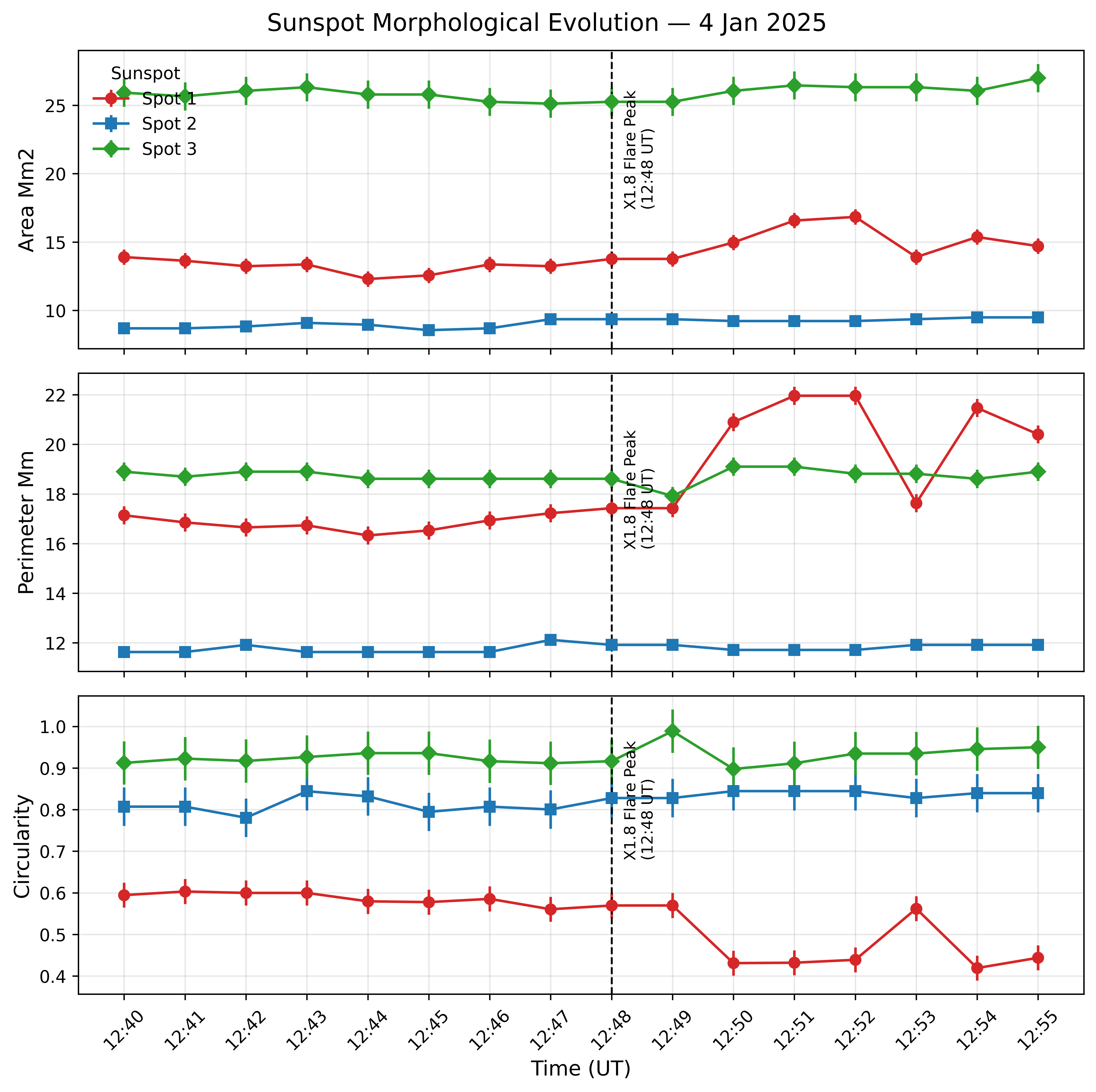}
        \caption{Sunspot Morphological Evolution}
        \label{fig:morphology_absolute}
    \end{subfigure}
    \hfill
    \begin{subfigure}[t]{0.48\textwidth}
        \centering
        \includegraphics[width=\linewidth]
        {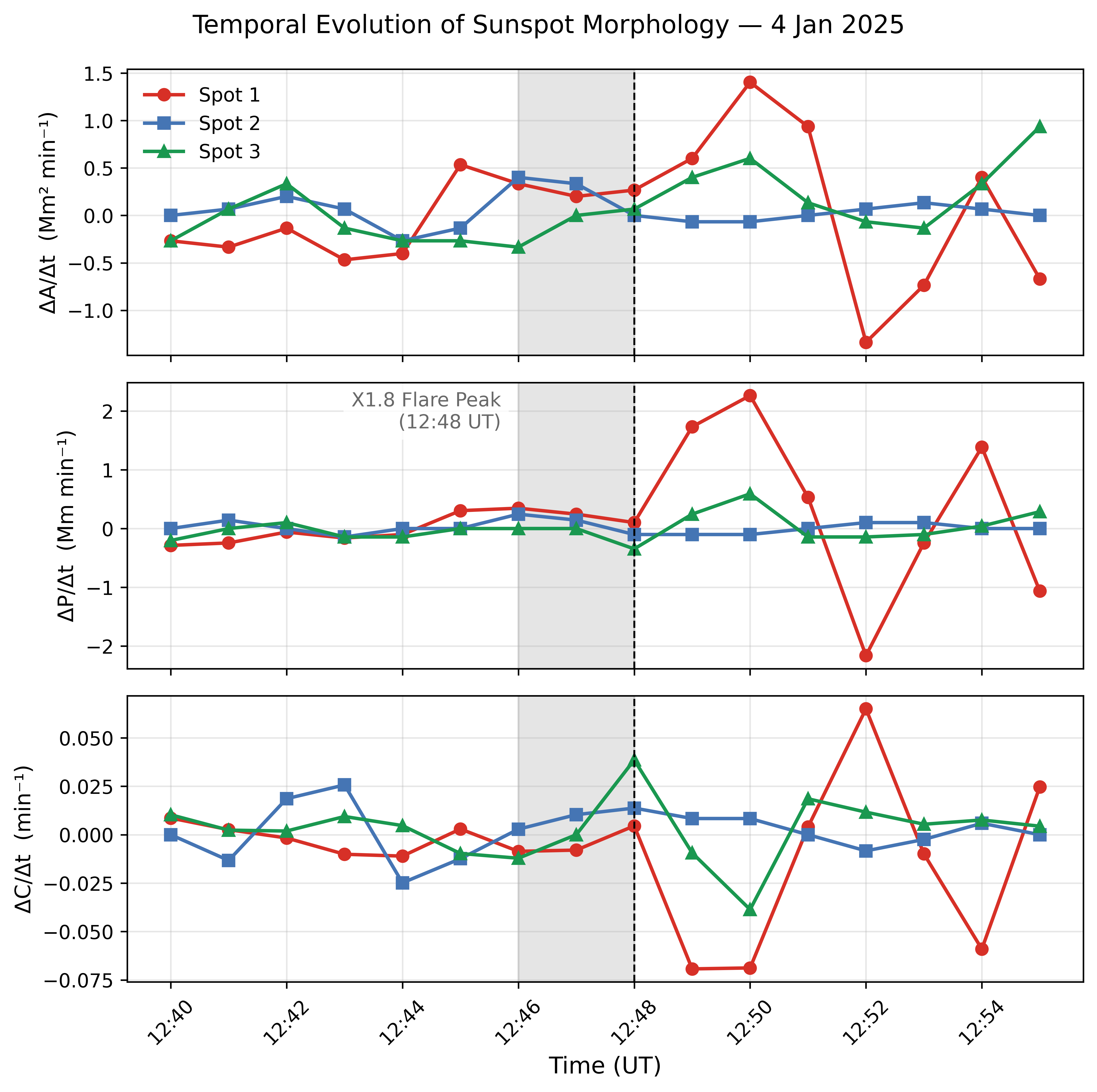}
        \caption{ Temporal Evolution of Sunspot Morphology }
        \label{fig:morphology_derivatives}
    \end{subfigure}
    \caption{(a)Shows the temporal evolution of the sunspot morphological parameters (area, perimeter \& circularity) for three sunspots inside AR 13947 during the eruption time interval. The peak flare time (12:48 UTC) is marked by vertical dashed line. (b) Associated time derivatives of the area $(\Delta A/\Delta t)$, perimeter $(\Delta P/\Delta t)$, \& circularity $(\Delta C/\Delta t)$, showing the localized and rapid changes, mainly for the Spot 1, in close temporal link with the flare peak.}
\end{figure}

\begin{table}[ht]
\centering
\caption{Detailed Sunspot Data with Uncertainties ($\pm \sigma$)}
\label{table:sunspot_data}
\adjustbox{max width=\textwidth}{%
\begin{tabular}{|c|c|c|c|c|@{\hspace{1.5em}}|c|c|c|c|c|}
\hline
\textbf{Time} & \textbf{S.} & \textbf{Area $\pm$ Unc.} & \textbf{Perim. $\pm$ Unc.} & \textbf{Circ. $\pm$ Unc.} & \textbf{Time} & \textbf{S.} & \textbf{Area $\pm$ Unc.} & \textbf{Perim. $\pm$ Unc.} & \textbf{Circ. $\pm$ Unc.} \\ \hline
12:40 & 1 & 13.90 $\pm$ 0.55 & 17.14 $\pm$ 0.34 & 0.5945 $\pm$ 0.033 & 12:47 & 3 & 25.13 $\pm$ 1.00 & 18.61 $\pm$ 0.37 & 0.9115 $\pm$ 0.051 \\
12:40 & 2 & 8.69 $\pm$ 0.34  & 11.63 $\pm$ 0.23 & 0.8071 $\pm$ 0.045 & 12:47 & 4 & 0.53 $\pm$ 0.02  & 1.46 $\pm$ 0.03  & 1.0000 $\pm$ 0.100 \\
12:40 & 3 & 25.93 $\pm$ 1.03 & 18.90 $\pm$ 0.38 & 0.9123 $\pm$ 0.051 & 12:48 & 1 & 13.77 $\pm$ 0.55 & 17.43 $\pm$ 0.35 & 0.5695 $\pm$ 0.032 \\
12:40 & 4 & 0.94 $\pm$ 0.04  & 2.50 $\pm$ 0.05  & 1.0000 $\pm$ 0.100 & 12:48 & 2 & 9.36 $\pm$ 0.37  & 11.92 $\pm$ 0.24 & 0.8278 $\pm$ 0.047 \\ \hline
12:41 & 1 & 13.63 $\pm$ 0.54 & 16.85 $\pm$ 0.33 & 0.6031 $\pm$ 0.034 & 12:48 & 3 & 25.26 $\pm$ 1.00 & 18.61 $\pm$ 0.37 & 0.9164 $\pm$ 0.051 \\
12:41 & 2 & 8.69 $\pm$ 0.34  & 11.63 $\pm$ 0.23 & 0.8071 $\pm$ 0.045 & 12:48 & 4 & 0.27 $\pm$ 0.01  & NaN              & 1.0000 $\pm$ 0.100 \\
12:41 & 3 & 25.66 $\pm$ 1.02 & 18.70 $\pm$ 0.37 & 0.9226 $\pm$ 0.052 & 12:49 & 1 & 13.77 $\pm$ 0.55 & 17.43 $\pm$ 0.35 & 0.5695 $\pm$ 0.032 \\
12:41 & 4 & 0.94 $\pm$ 0.04  & 2.50 $\pm$ 0.05  & 1.0000 $\pm$ 0.100 & 12:49 & 2 & 9.36 $\pm$ 0.37  & 11.92 $\pm$ 0.24 & 0.8278 $\pm$ 0.047 \\ \hline
12:42 & 1 & 13.23 $\pm$ 0.53 & 16.65 $\pm$ 0.33 & 0.5997 $\pm$ 0.034 & 12:49 & 3 & 25.26 $\pm$ 1.00 & 17.92 $\pm$ 0.36 & 0.9887 $\pm$ 0.056 \\
12:42 & 2 & 8.82 $\pm$ 0.35  & 11.92 $\pm$ 0.24 & 0.7805 $\pm$ 0.044 & 12:49 & 4 & 0.27 $\pm$ 0.01  & NaN              & 1.0000 $\pm$ 0.100 \\
12:42 & 3 & 26.06 $\pm$ 1.03 & 18.90 $\pm$ 0.38 & 0.9170 $\pm$ 0.052 & 12:50 & 1 & 14.97 $\pm$ 0.59 & 20.89 $\pm$ 0.41 & 0.4309 $\pm$ 0.024 \\
12:42 & 4 & 0.80 $\pm$ 0.03  & 2.19 $\pm$ 0.04  & 1.0000 $\pm$ 0.100 & 12:50 & 2 & 9.22 $\pm$ 0.37  & 11.71 $\pm$ 0.23 & 0.8445 $\pm$ 0.047 \\ \hline
12:43 & 1 & 13.36 $\pm$ 0.53 & 16.73 $\pm$ 0.33 & 0.5997 $\pm$ 0.034 & 12:50 & 3 & 26.06 $\pm$ 1.03 & 19.10 $\pm$ 0.38 & 0.8976 $\pm$ 0.050 \\
12:43 & 2 & 9.09 $\pm$ 0.36  & 11.63 $\pm$ 0.23 & 0.8443 $\pm$ 0.047 & 12:51 & 1 & 16.57 $\pm$ 0.66 & 21.96 $\pm$ 0.44 & 0.4319 $\pm$ 0.024 \\
12:43 & 3 & 26.33 $\pm$ 1.04 & 18.90 $\pm$ 0.38 & 0.9264 $\pm$ 0.052 & 12:51 & 2 & 9.22 $\pm$ 0.37  & 11.71 $\pm$ 0.23 & 0.8445 $\pm$ 0.047 \\
12:43 & 4 & 0.80 $\pm$ 0.03  & 2.19 $\pm$ 0.04  & 1.0000 $\pm$ 0.100 & 12:51 & 3 & 26.46 $\pm$ 1.05 & 19.10 $\pm$ 0.38 & 0.9114 $\pm$ 0.051 \\ \hline
12:44 & 1 & 12.30 $\pm$ 0.49 & 16.33 $\pm$ 0.32 & 0.5795 $\pm$ 0.033 & 12:52 & 1 & 16.84 $\pm$ 0.67 & 21.96 $\pm$ 0.44 & 0.4389 $\pm$ 0.025 \\
12:44 & 2 & 8.95 $\pm$ 0.36  & 11.63 $\pm$ 0.23 & 0.8319 $\pm$ 0.047 & 12:52 & 2 & 9.22 $\pm$ 0.37  & 11.71 $\pm$ 0.23 & 0.8445 $\pm$ 0.047 \\
12:44 & 3 & 25.79 $\pm$ 1.02 & 18.61 $\pm$ 0.37 & 0.9358 $\pm$ 0.053 & 12:52 & 3 & 26.33 $\pm$ 1.04 & 18.81 $\pm$ 0.37 & 0.9347 $\pm$ 0.053 \\
12:44 & 4 & 1.07 $\pm$ 0.04  & 2.92 $\pm$ 0.06  & 1.0000 $\pm$ 0.100 & 12:53 & 1 & 13.90 $\pm$ 0.55 & 17.63 $\pm$ 0.35 & 0.5619 $\pm$ 0.032 \\ \hline
12:45 & 1 & 12.56 $\pm$ 0.50 & 16.53 $\pm$ 0.33 & 0.5776 $\pm$ 0.032 & 12:53 & 2 & 9.36 $\pm$ 0.37  & 11.92 $\pm$ 0.24 & 0.8278 $\pm$ 0.047 \\
12:45 & 2 & 8.55 $\pm$ 0.34  & 11.63 $\pm$ 0.23 & 0.7946 $\pm$ 0.045 & 12:53 & 3 & 26.33 $\pm$ 1.04 & 18.81 $\pm$ 0.37 & 0.9347 $\pm$ 0.053 \\
12:45 & 3 & 25.79 $\pm$ 1.02 & 18.61 $\pm$ 0.37 & 0.9358 $\pm$ 0.053 & 12:53 & 4 & 0.27 $\pm$ 0.01  & NaN              & 1.0000 $\pm$ 0.100 \\
12:45 & 4 & 0.94 $\pm$ 0.04  & 2.63 $\pm$ 0.05  & 1.0000 $\pm$ 0.100 & 12:54 & 1 & 15.37 $\pm$ 0.61 & 21.47 $\pm$ 0.43 & 0.4191 $\pm$ 0.024 \\ \hline
12:46 & 1 & 13.36 $\pm$ 0.53 & 16.94 $\pm$ 0.34 & 0.5854 $\pm$ 0.033 & 12:54 & 2 & 9.49 $\pm$ 0.38  & 11.92 $\pm$ 0.24 & 0.8396 $\pm$ 0.047 \\
12:46 & 2 & 8.69 $\pm$ 0.34  & 11.63 $\pm$ 0.23 & 0.8071 $\pm$ 0.045 & 12:54 & 3 & 26.06 $\pm$ 1.03 & 18.61 $\pm$ 0.37 & 0.9455 $\pm$ 0.053 \\
12:46 & 3 & 25.26 $\pm$ 1.00 & 18.61 $\pm$ 0.37 & 0.9164 $\pm$ 0.051 & 12:55 & 1 & 14.70 $\pm$ 0.58 & 20.40 $\pm$ 0.40 & 0.4438 $\pm$ 0.025 \\
12:46 & 4 & 0.53 $\pm$ 0.02  & 1.46 $\pm$ 0.03  & 1.0000 $\pm$ 0.100 & 12:55 & 2 & 9.49 $\pm$ 0.38  & 11.92 $\pm$ 0.24 & 0.8396 $\pm$ 0.047 \\ \hline
12:47 & 1 & 13.23 $\pm$ 0.53 & 17.22 $\pm$ 0.34 & 0.5604 $\pm$ 0.031 & 12:55 & 3 & 27.00 $\pm$ 1.07 & 18.90 $\pm$ 0.38 & 0.9499 $\pm$ 0.053 \\
12:47 & 2 & 9.36 $\pm$ 0.37  & 12.12 $\pm$ 0.24 & 0.8003 $\pm$ 0.045 &       &   &                  &                  &                    \\ \hline
\end{tabular}%
}
\end{table}
The photospheric response of the AR 13947 can be observed by investigating the morphological evolution of the individual sunspots present within the active region.
As shown in Figure~\ref{fig:morphology_absolute} \& detailed in Table~\ref{table:sunspot_data}, Sunspot 1(red) shows the most pronounced \& abrupt changes in its morphological structure. Prior to the flare at 12:48 UTC, the Sunspot 1 maintained a relatively stable area of Approx. 13.5 $\mathrm{Mm}^2$
 \& a circularity of 0.60.
Within minutes following the flare maximum, the Sunspot 1 experienced a rapid expansion, with its area increasing to 16.84±0.67 $\mathrm{Mm}^2$ \& its perimeter increases from 17.43 to 21.96 Mm by 12:52 UTC.
Further this behavior is highlighted in the temporal derivate plots (see Fig.~\ref{fig:morphology_derivatives}) where $\frac{\Delta A}{\Delta t}$ and $\frac{\Delta P}{\Delta t}$ for Sunspot 1 shows a strong positive excursion instantly after the flare peak.
Most notably, between 12:49 \& 12:50 UTC the circularity of Sunspot 1 decreased sharply from 0.57 to 0.43.
The rapid increase in perimeter \& decrease in circularity suggests that the sunspot did not simply grow in size, but also underwent deformation, further becoming more irregular and elongated \cite{Deng2005}.
Such behavior is consistent with a rapid response by photosphere to coronal magnetic restructuring, possibly associated with abrupt changes in the Lorentz forces serving on the lower atmosphere \cite{Fisher2012}.
In contrast, the Sunspot 2 (blue) and Sunspot 3 (green) remained relatively stable over the entire event, with Sunspot 3 sustaining a high circularity ($>$0.90) \& only some minor fluctuations in area.
This localized morphological response confirms that the strongest photospheric reconfiguration was concentrated specifically around the magnetic core of Sunspot 1.
A much smaller Sunspot 4 was also present within the AR 13947; however, its morphological parameters could not be
Robustly quantified due to notable pixelation \& segmentation uncertainties associated from its small spatial extent. As a result, this feature was excluded from the analysis to avoid non-physical trends.
\subsection{Eruption Physics: Non-Thermal Particle Acceleration (STIX)}
\begin{figure}[h!]
    \centering
    \includegraphics[width=\textwidth]{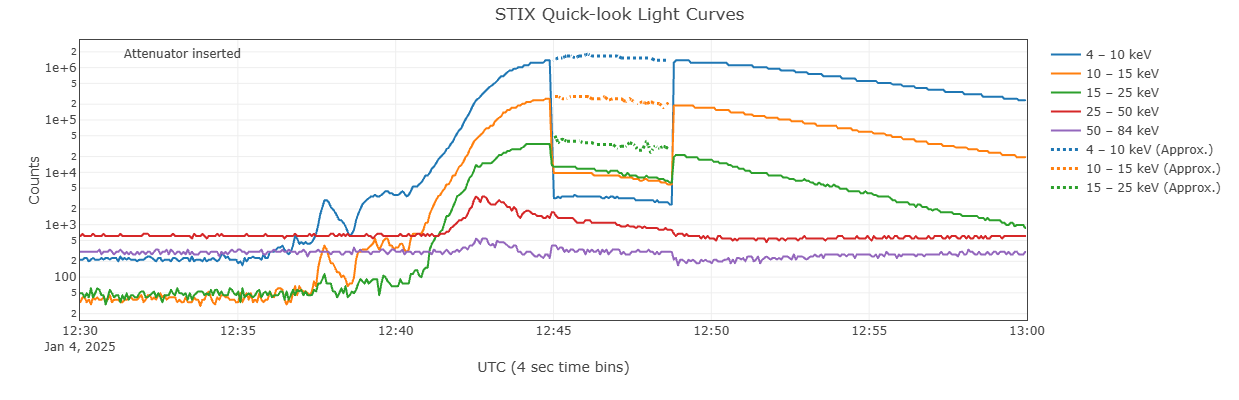}
    \caption{The STIX instrument onboard \textit{Solar Orbiter} observed the hard X-ray light curves of the X1.8-class flare. The multi-energy profiles (1--84~keV) display a pronounced impulsive phase, with enhanced high-energy emission ($\geq 25$~keV) preceding the thermal peak in the lower-energy channels. The vertical feature associated with attenuator insertion is clearly identified. The temporal ordering of the high- and low-energy emissions is consistent with non-thermal electron acceleration followed by thermal plasma heating, in agreement with the Neupert effect.}
    \label{fig:STIX}
\end{figure}
To study the non-thermal particle acceleration processes during the X1.8-class flare, we analyzed hard X-ray (HXR) observations from the STIX instrument onboard \textit{Solar Orbiter} (see Fig.~\ref{fig:STIX}).
The multi-energy light curves provide a timing-based comparison between the impulsive non-thermal emission and the subsequent thermal plasma response, while accounting for the instrumental attenuator transitions visible in the data \cite{Brown1971}.
The impulsive phase of the flare began at approximately 12:41~UTC and is characterized by a rapid increase in count rates across all energy bands.
Notably, the high energy channels- particularly 25-50 keV (red) \& the 50-84 keV (purple) bands, reach peak emission between 12:43 and 12:45 UTC, preceding the thermal peak observed in the lower-energy (4-15 keV) channels and GOES X-ray data.
This temporal offset is consistent with the Neupert-type behavior, suggesting that the energy released via magnetic reconnection initially accelerated the non-thermal electrons \cite{Neupert1968}.
These accelerated electrons generate HXR emission via bremsstrahlung upon impact with the dense chromosphere while concurrently depositing energy that heated the ambient plasma, leading to ‘chromospheric evaporation’ that populates the flare loop with the hot, soft X-ray emitting plasma observed in the lower energy channels \cite{Dennis1993}.
The detection of significant emission off up to 84 keV confirms that the AR 13947 was a highly efficient particle accelerator.
Additionally, small pre-flare bursts observed between 12:37 and 12:40 in the 54–10 keV energy range suggest localized pre-heating or precursor magnetic reconnection processes that destabilized the magnetic configuration preceding the main eruption.

\subsection{Large-Scale Consequences: CME and Coronal Shock (e-CALLISTO)}

 \subsubsection{Coronal Dimming}
\begin{figure}[H]
    \centering
    \begin{subfigure}[t]{0.48\textwidth}
        \centering
        \includegraphics[width=\linewidth]{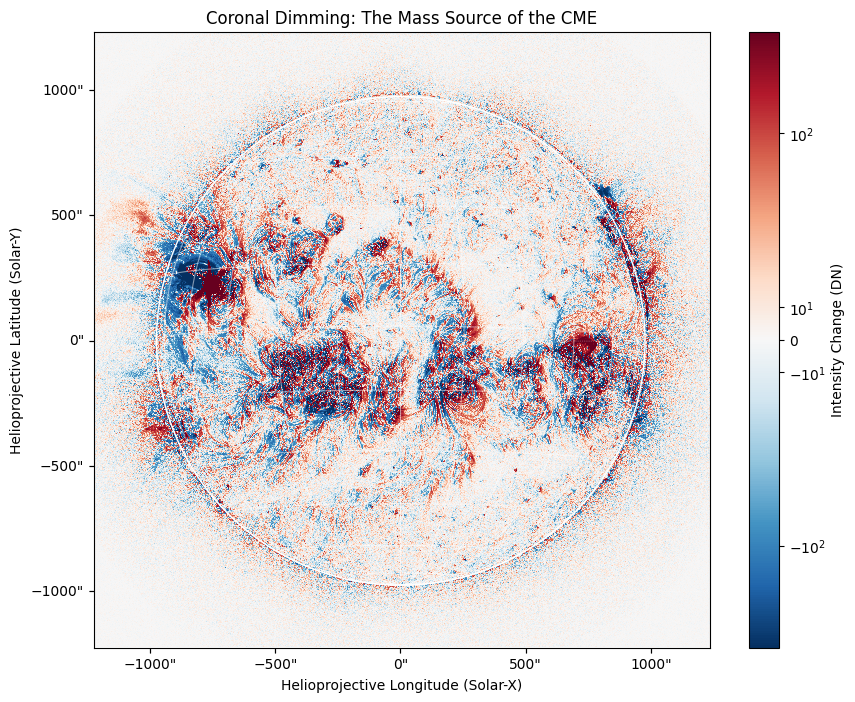}
        \caption{Full-disk EUV base-difference map.}
        \label{fig:dimming_full}
    \end{subfigure}
    \hfill
    \begin{subfigure}[t]{0.48\textwidth}
        \centering
        \includegraphics[width=\linewidth]{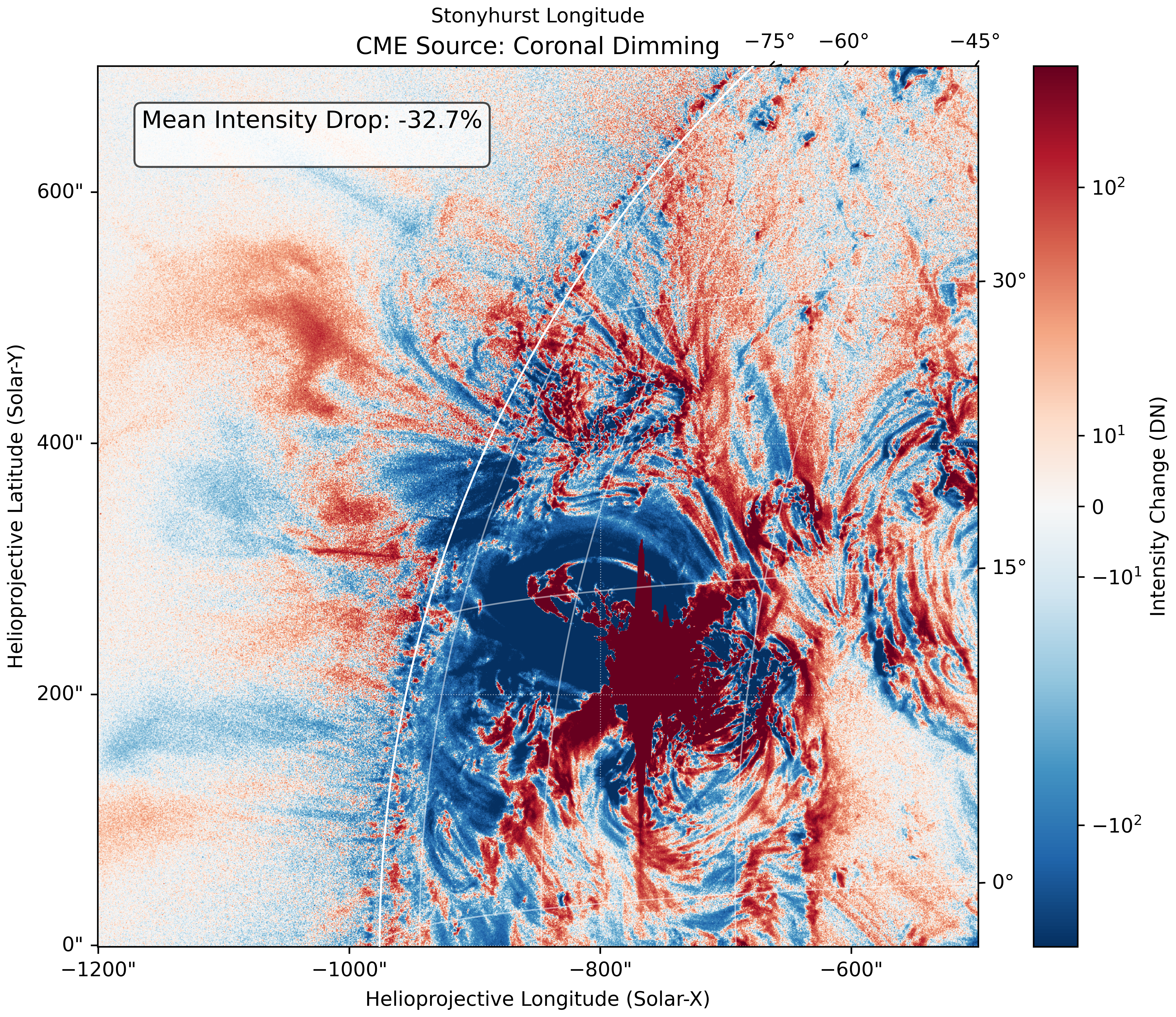}
        \caption{Close-up EUV base-difference image showing localized coronal dimming near AR 13947.}
        \label{fig:dimming_closeup}
    \end{subfigure}
    \caption{Coronal dimming associated with the X1.8-class flare, illustrated using EUV base-difference images. Blue regions represent intensity depletion, indicative of plasma evacuation during the CME, while red regions denote enhanced emission. The dimming is predominantly concentrated around Active Region 13947 in the northwestern quadrant of the solar disk. The mean intensity decrease in the dimming region is approximately $-32.7\%$.}
    \label{fig:Overalldimming}
\end{figure}

 To explore the large-scale eruptive effects of the X1.8 flare, we analysed coronal dimming signatures using the EUV base-difference images (see Fig.~\ref{fig:Overalldimming})
The transient depletion in the density of solar corona is represented by the Coronal dimmings, also it serves a robust indirect proxy for the evacuation of plasma during a CME \cite{Dissauer2018}.
The Full-disk difference map (Fig.~\ref{fig:dimming_full}) shows widespread intensity depletion (depicted in blue) predominately concentrated in the north-western quadrant surrounding the AR13947, although the effects extend substantially across the solar disk.
Furthermore a zoomed-in view of the CME source region (Fig.~\ref{fig:dimming_closeup}) identifies mean intensity drop of -32.7\% within the dimming core. This evacuation of material supports that the X1.8 class flare was a fully eruptive event.\\
The deep, spatially localized dimming region is consistent with a principal mass contribution to the CME, consistent with the associated large-scale reconfiguration of the magnetic field lines during the flare eruption \cite{Zhukov2004}.
The spatial correlation between this dimming \& the subsequent white -light CME observed by SOHO/LASCO strongly supports the interpretation that the erupting magnetic structure originated from the core of AR 13947.

\subsubsection{Type II Radio Burst}
\label{Section_radio_burst}
\begin{figure}[H]
    \centering
    \includegraphics[width=0.95\textwidth]{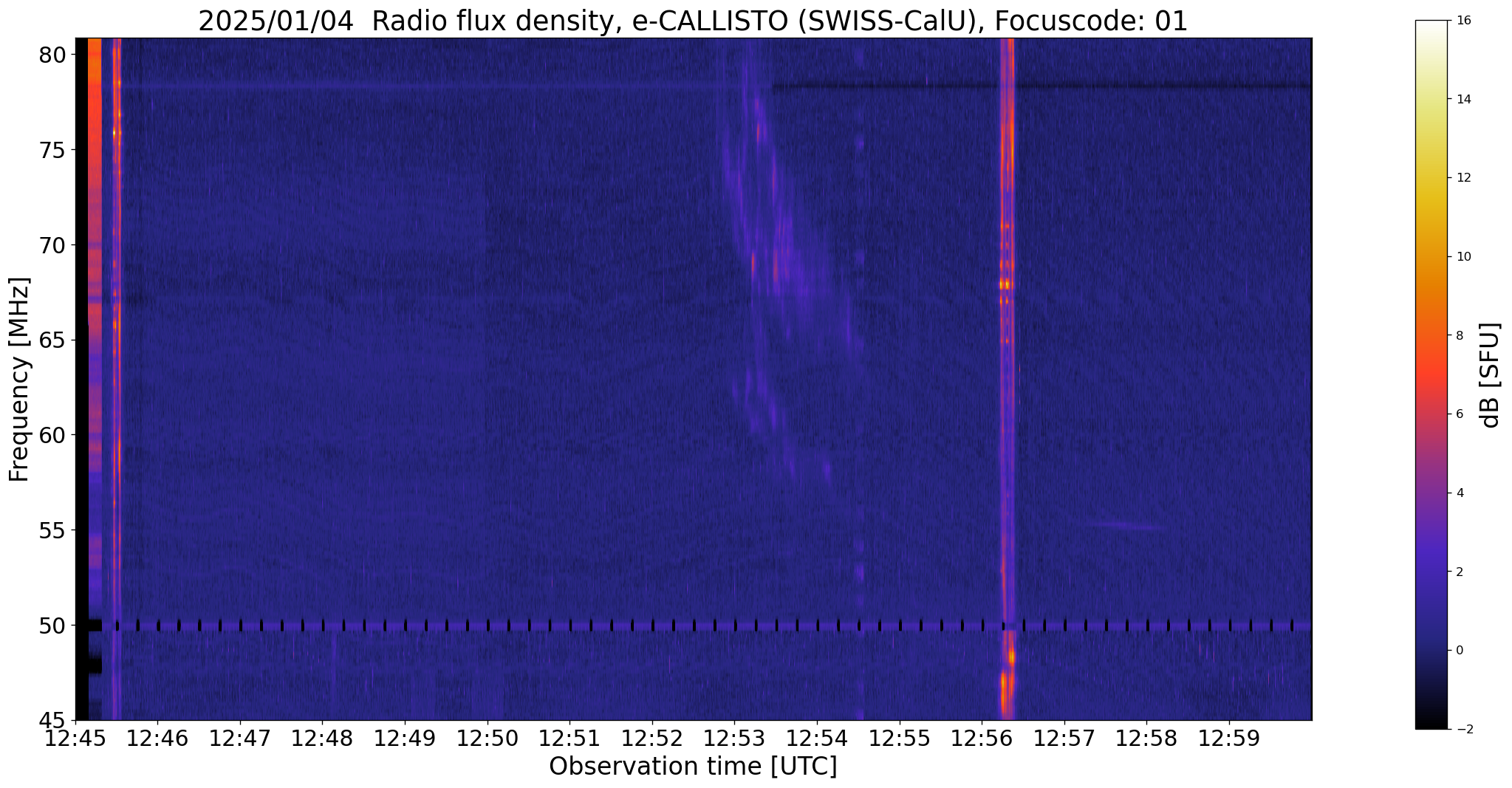}
    \caption{Shock-driven radio spectrum detected by the e-CALLISTO spectrometer (SWISS-CAlU station) during the eruption time interval. A slow drifting emission signature in the 60-70 MHz range is visible between time range of 12:53 \& 12:54 UTC, signature of a Type II radio burst linked with a coronal shock.}
    \label{fig:ecallisto_dynamic_spectrum}
\end{figure}

Subsequent to the impulsive phase of X1.8 flare, a metric Type II radio burst was detected by the e-CALLISTO spectrometer (SWISS-CAIU station) during the interval 12:53 \& 12:54 UTC (see Fig.~\ref{fig:ecallisto_dynamic_spectrum}).
We identified a burst exhibiting slow-drift emission feature which is moving from higher to lower frequencies, a diagnostic signature of a magnetohydrodynamics (MHD) shock wave propagation radially outward via solar corona \cite{Nelson1985}.
From approximately 75 MHz to 60 MHz over a 60 second time interval, the emission was observed drifting. Such bursts are generally associated with the piston-driven shock wave front prior Coronal Mass Ejection (CME)\cite{Gopalswamy2001}.
Identification of this Type II radio burst provides observational evidence of the flare’s eruptive nature, also allows for the estimation of the shock’s propagation speed.

\subsubsection{Shock Speed Estimation}
To measure the dynamics of the eruption, we estimated the coronal shock speed ($V$) from the frequency drift rate $\left(\frac{df}{dt}\right)$ of Type II radio burst detected in the e-CALLISTO spectrum (Fig.~\ref{fig:ecallisto_dynamic_spectrum}). Given the flare’s origin in the magnetically complex and dense AR 13947, we adopted a Newkirk-type active-region density model to explain the intensified coronal density and steeper gradients \cite{Newkirk1961,Mann1999}.
For the calculations, we utilized a coronal density scale height of $L_n \approx 0.11\,R_\odot \; (\sim 75{,}000\,\mathrm{km})$ was assumed, which is consistent with the lower-coronal active-region environments.
The shock speed is calculated using the following relation:
\begin{equation}\label{speed}
V = \frac{2 \cdot L_n}{f} \cdot \left| \frac{df}{dt} \right|
\end{equation}
From the e-CALLISTO data, we obtained the  parameters: frequency change $(\Delta f)$: 75 MHz$\rightarrow$
60 MHz $\Rightarrow$ $\mathrm{d}f$ = $15\,\mathrm{MHz}$, time duration $(\Delta t)$: 60 s, drift Rate $\left(\frac{df}{dt}\right)$: 15/60= $0.25\,\mathrm{MHz\,s^{-1}}$, and center Frequency (f): (75+60)/2= $67.5\,\mathrm{MHz}$.
Substituting these values into the given equation (\ref{speed}), we calculated the shock speed.
The estimated shock speed $\approx 556\,\mathrm{km\,s^{-1}}$ is consistent with a moderately fast coronal shock capable of generating a Type II burst in lower corona region. Combined with the primarily northwestern propagation direction inferred from LASCO observations (Section~\ref{CME}), this intermediate shock speed indicates that the eruption was unlikely to produce a strong geoeffective impact at the Earth, either due to a non-Earth directed trajectory or shock weakening with the height.

\subsubsection{CME Trajectory}
\label{CME}

\begin{figure}[H]
    \centering
    \includegraphics[width=0.85\textwidth]{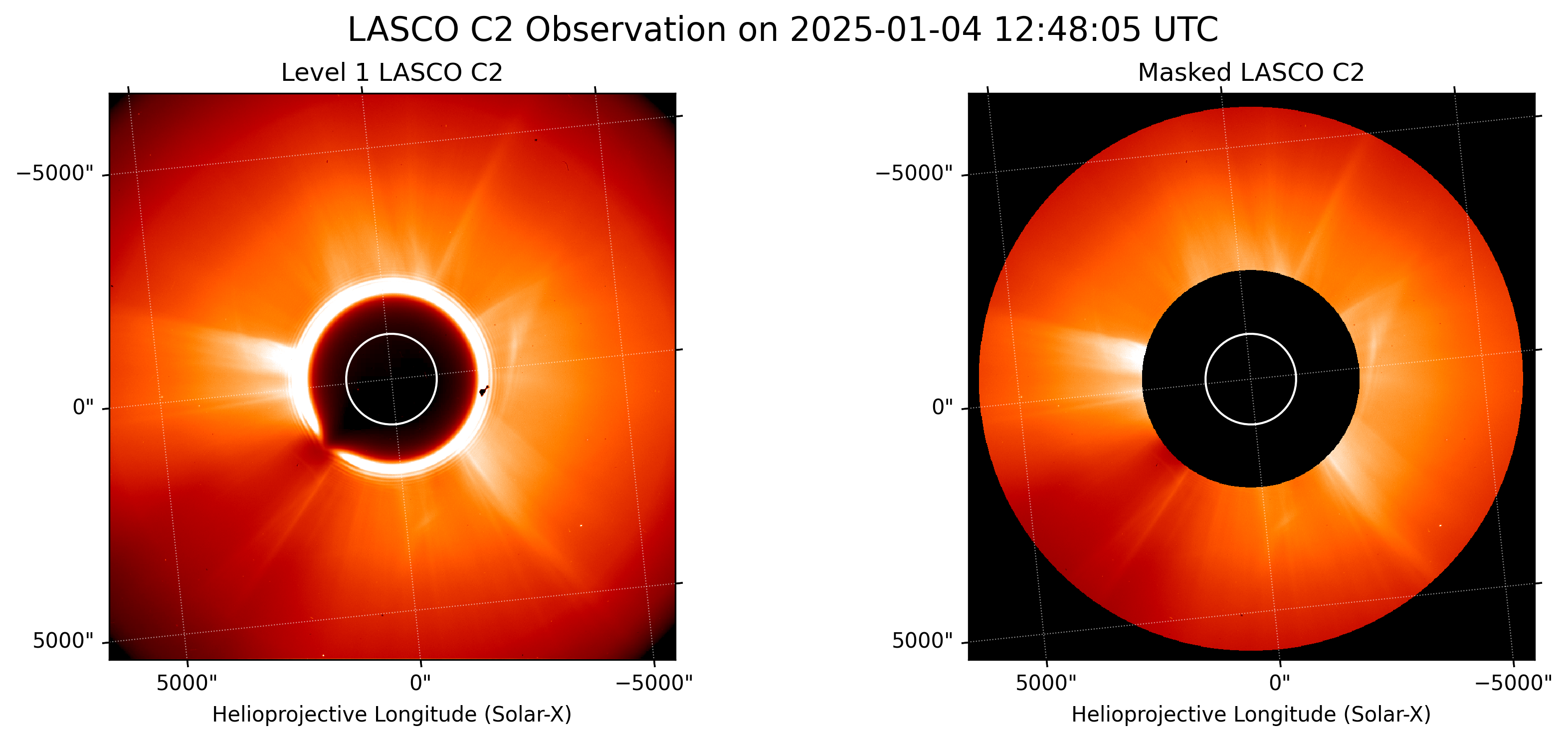}
    \caption{(a)Shows the large-scale coronal arrangement throughout the X1.8 flare period, captured at 12:48:05 UTC by SOHO/LASCO C2 level-1 coronagraph image. (b) Associated masked LASCO C2 image with occulting disk \& peripheral background excluded to enhance the clarity of outward moving coronal signatures linked with the CME.}
    \label{fig:lasco_c2_soho}
\end{figure}

As we know, solar flare is outlined by its intense outburst of electromagnetic radiation and events like the $X$1.8 class solar flare are typically connected with  CME, an eruptive, billion-ton cloud of plasma released into the heliospheric region
To verify and study this eruption, we employ chronagraph data from the Large Angle \& Spectrometric Coronagraph (LASCO) C2 instrument, on board the Solar and Heliospheric Observatory (SOHO).
The LASCO C2 instrument uses an occulting disk to obstruct the direct sunlight and further monitors the outer solar corona about 1.5 to 6 solar radii, which makes it a leading tool for analyzing and tracing the CMEs as they move away from the Su.\cite{Brueckner1995}.
Fig.~(\ref{fig:lasco_c2_soho}) shows the LASCO C2 image pair, captured at 12:48:05 UTC, which presents a definite proof of the associated CME with the flare. The high-cadence correlation, merged with the eruption’s origin from the northwestern limb, which is the active region 13947, validates a well-defined link between the flare and the CME.
The left panel of the Fig.~(\ref{fig:lasco_c2_soho}), labeled Level 1 'LASCO C2', shows the raw observational data, which reveal a luminous multi-stratified cloud of plasma, extending rapidly into the field of view, visibly distinct from the quiescent.
The processed image on the right of Fig.~(\ref{fig:lasco_c2_soho}) 'Marked LASCO C2', which is a differenced image that subtracts a prevent background and also separates the transient CME, revealing the internal structure with significant clarity.
The most essential property of this CME for space weather forecasting is the trajectory, which can be estimated by its point of origin on solar disk \cite{Gopalswamy2009}.
Emanating from the northwestern limb of the Sun (AR 13947). The CME was also launched into the northwestern quadrant of the heliosphere, majorly away from the Sun-Earth line. Its path ensures that the plasma with the enclosed magnetic field was passed to the side of our planet, missing Earth thoroughly. Thus, there was no major geomagnetic storm, or disruptions to the technologies from this event. This event provides a standard example of how the solar active region’s location is as significant as its inherent activity in calculating its ultimate effect on the terrestrial space environment.

\section{Discussion and Conclusion}\label{sect4}
This work presents a detailed source-to-eruption analysis of the X1.8 solar flare linked with a CME produced by Active region 13947 on 4 January 2025.
By integrating the data from ground \& space-based observatories, we have reconstructed the ordered sequence of this event, from its magnetic source to its effects in heliosphere \cite{Shibata2011}.
The photospheric magnetic activities indicates that the solar flare was triggered by impulsive magnetic shearing \& stored free energy around the complex AR 13947.
The morphological evolution tracking of sunspots showed a high-cadence variation of the eruption; mainly the deformation and expansion of Sunspot 1 (as indicated by a $25\%$ decrease in circularity) illustrates observational evidence of Lorentz force’s impact on the lower atmospheric region during the magnetic reconnection process \cite{Fletcher2011}.
The discharge of energy followed the standard solar flare model, validation by the temporal off-set between the hard (HRX-STIX) non-thermal peaked \& the Soft X-ray (SXR-GOES) thermal maximum, which shows the classical signature of Neupert effect.
The SDO/AIA \& the NUV observations by Aditya L1/SUIT revealed the multi-thermal response that confirms the energy deposition from chromosphere to the coronal region, with the 131 \AA\ channel effectively isolating the super-heated reconnection core ($>10~\mathrm{MK}$) .
The impulsive nature of the event was confirmed by a Type II radio burst detection \& coronal dimming of around $-32.7\%$. The computed shock speed is $\approx 556~\mathrm{km\,s^{-1}}$  making this moderately fast, eruptive event. Although, despite high X-ray intensity of the flare, the associated CME was non-geoeffective.
Our analysis of LASCO C2 trajectory and solar coordinates (north-western limb) validates that the plasma was oriented away from the Sun-Earth line.
In conclusion, the Active Region 13947 shows a highly productive region. This work emphasizes the necessity of multi-instrument datasets in understanding these types of eruptive events in prior.

\section*{CRediT authorship contribution statement}

Akash Vinod Shirke: Methodology, Software, Writing -- original draft.  

Sakshi Sheshrao Charde: Conceptualization, Software, Writing -- original draft.  

Balendra Pratap Singh: Supervision, Software, Visualization, Writing -- review \& editing.

\section*{Data Availability}
All data supporting the findings of this study are publicly available from the official archives of GOES, SDO (AIA and HMI), Solar Orbiter (STIX), e-CALLISTO, Aditya-L1 (SUIT), and SOHO.

\section*{Acknowledgments}
The authors acknowledge the use of the SunPy open-source Python package. Balendra P.~Singh gratefully acknowledges the support from UPES through the project code UPES/R\&D-SOAE/25062025/23.

\end{document}